\documentclass[11pt]{article}
\usepackage[defaultcolor=black]{changes} % for revision
\usepackage{hyperref}
\usepackage{amsmath, amsfonts, amssymb, mathrsfs}
\usepackage{dcolumn}
\usepackage{caption}
\usepackage{subcaption}
\usepackage{filemod}
\usepackage{natbib}
\usepackage[ruled, vlined]{algorithm2e}
\usepackage{floatrow}
\usepackage{setspace}
\usepackage{verbatim}
\usepackage{graphicx}
\usepackage{bbm}
\usepackage{bm}

\hypersetup{
    colorlinks=true,
    linkcolor=blue,%magenta, 
    urlcolor=black,
    citecolor=blue,%magenta, 
    }

\title{\vspace{-1cm} \added{Gradient-enhancement and Gradient Predictions for Deep
Gaussian Process Modeling of Expensive Computer Experiments}}
\author{Annie S. Booth\thanks{Corresponding author: Department of Statistics, 
	Virginia Tech, {\tt annie\_booth@vt.edu}}}
\date{\today}
\begin{document}

\maketitle

\begin{abstract} 
Deep Gaussian processes (DGPs) are popular surrogate models for complex nonstationary computer
experiments.  DGPs use one or more latent Gaussian processes (GPs) to warp the input
space into a plausibly stationary regime, then use typical GP regression
on the warped domain.  While this composition of GPs is conceptually 
straightforward, the functional nature of the
multi-dimensional latent warping makes Bayesian posterior inference challenging.
Traditional GPs with smooth kernels are naturally suited for the integration of gradient
information, but the integration of gradients within a DGP
presents new challenges and has yet to be explored.  We propose a novel and comprehensive
Bayesian framework for DGPs with gradients that facilitates both gradient-enhancement
and gradient posterior predictive distributions.  \added{Our focus is on surrogate
modeling of expensive, deterministic, and nonstationary computer experiments.
Gradient-enhancement is most impactful when data is limited, and
gradient predictions are most useful for downstream surrogate modeling tasks
like optimization and active learning.  We benchmark both contributions 
(gradient-enhanced DGPs and DGP gradient predictions), separately and together,
on a variety of nonstationary test functions as well as real quantum
mechanics computer experiments that simulate molecular energy and forces as a function 
of atomic position.  On nonstationary surfaces, our gradient-enhanced DGPs outperform 
gradient-enhanced GPs and non-enhanced DGPs, and our DGP gradient predictions are 
more effective than GP gradient predictions.
We provide open-source software in the ``deepgp'' package on CRAN, with optional 
Vecchia approximation to circumvent cubic computational bottlenecks.}
\end{abstract}

\added{gradient-enhanced kriging}, surrogate, Vecchia approximation, 
uncertainty quantification

%%%%%%%%%%%%%%%%%%%%%%%%%%%%%%%%%%%%%%%%%%%%%%%%%%%%%%%%%%%%%%%%%%%%%
\section{Introduction}\label{sec:intro}
%%%%%%%%%%%%%%%%%%%%%%%%%%%%%%%%%%%%%%%%%%%%%%%%%%%%%%%%%%%%%%%%%%%%%

Complex computer simulation experiments, whose computational expense
may restrict evaluation budgets, require effective surrogate models
(or ``emulators'') that may stand in place of true simulation evaluations at unobserved 
inputs.  A good surrogate model should provide accurate predictions with 
appropriate uncertainty quantification 
(UQ) at a computational cost significantly less than that of the true simulator
\citep{santner2003design,gramacy2020surrogates}. 
It should also: (i) effectively leverage all available training data
and (ii) facilitate downstream tasks such as active learning, 
calibration, and optimization.  While deep Gaussian processes 
\citep[DGPs;][]{damianou2013deep} have been gaining traction as surrogates
for nonstationary computer experiments
\citep{sauer2023deep}, there are still situations where they fall short on
these latter two tasks.  \added{In this work we advance DGPs as surrogates 
by incorporating gradients in two ways.  First, we use gradient-enhancement
to improve DGP predictions by conditioning on observed gradients.
Second, we obtain posterior predictions of the DGP's gradient
(which may or may not be gradient-enhanced).}
Upgrading the DGP to enable
gradient-enhancement will facilitate task (i) when the computer simulation
is able to return gradient information, which is common
in physics and engineering applications thanks to adjoint solvers
\citep[e.g.,][]{othmer2014adjoint,wang2015application,jacobson2021adjoint}.
Upgrading the DGP to return
predictions of the gradient at unobserved inputs will facilitate
task (ii), particularly if the downstream task requires gradient-based
numerical optimization of an acquisition function.

\added{We are motivated by expensive, deterministic, and nonstationary 
computer experiments whose complex response surfaces warrant the flexibility of
a DGP surrogate.  When evaluations are computationally expensive 
and data sizes are limited, gradient-enhancement can have a huge impact on the
predictive accuracy of the surrogate.  Additionally, strategic sequential designs
may be necessary to maximize learning from a tight evaluation budget
\citep[often called ``active learning,'' e.g.,][]{cohn1994active,song2026efficient},
which may warrant gradient-based optimization of acquisition functions.
For example, computer experiments of quantum mechanics that simulate molecular
energy and forces as a function of a molecule's atomic structure can be 
necessarily complex and computationally expensive 
\citep{buluta2009quantum,georgescu2014quantum}.  Recent efforts have sought
to compile massive public repositories of these simulation results
\citep[e.g.,][]{gabellini2024openqdc},
yet the myriad of potential molecules still limits the number of observations
that is practically attainable for each specific molecule.  Gradients are
particularly relevant for this application area as a molecule's nuclear forces
represent the negative gradient of its potential energy.  We will revisit this
application in Section \ref{sec:quantum}, where we consider molecules with 
25 or fewer available observations.}

The incorporation of gradients within typical Gaussian processes (GPs) is 
relatively straightforward \citep{rasmussen2006gaussian}, with early works dating back 
to \citet{morris1993bayesian}.  Closed-form predictions of GP gradients 
have been used for a variety of tasks including Bayesian optimization with expected
improvement \citep[e.g.,][]{ament2023unexpected}, dimension reduction 
\citep[e.g.,][]{wycoff2021sequential}, and sensitivity analysis 
\citep[e.g.,][]{wycoff2021gradient}.  Gradient-enhanced GP regression (also 
known as ``gradient-enhanced kriging'') has been thoroughly explored 
\citep[e.g.,][]{solak2002derivative,dwight2009efficient,de2014improvements,bouhlel2019gradient}
with applications to optimization \citep{wu2017bayesian,kaappa2021global} and 
multifidelity modeling \citep{ulaganathan2015performance,deng2020multifidelity}.
Gradient-enhancement can be particularly challenging as it opens the door to 
ill-conditioned covariance matrices \citep{dalbey2013efficient,he2018instability,marchildon2023non}
and computational bottlenecks \citep{eriksson2018scaling}.

Despite their popularity, traditional GP surrogates are ill-equipped to handle
some of the complex response surfaces that are prevalent in modern computer simulations.
The stationarity of the GP covariance kernel forces it to impart the same correlation
structure across the entire domain, which is at odds with simulations that experience
shifting dynamics \citep[say, when a jet engine ignites,][]{stumbar2025validated}.
While there are a variety of nonstationary GP adaptations aimed at improving surrogate
flexibility while retaining fidelity \citep{booth2024non}, deep Gaussian processes 
have risen to the top.  Originating in spatial statistics \citep{schmidt2003bayesian}
and popularized for machine learning \citep{damianou2013deep,bui2016deep},
DGPs have shown great promise as surrogates for nonstationary computer
simulations \citep{rajaram2021empirical,marmin2022deep,ming2023deep,yazdi2024deep}.

DGPs operate through functional compositions of GPs.  Although deeper
variations exist \citep{dunlop2018deep}, we will restrict our work here
to those with two GP layers, resulting in a single latent space.  Additional
depth \added{may come} at significant computational expense and is often unwarranted
for surrogate modeling tasks \citep{sauer2023active}.  The inner GP layer defines a 
warping of the original inputs, and the outer GP layer acts as the primary regression model.
\added{Inferring an appropriate latent warping can be challenging due to its functional and
multi-dimensional nature.}  Many works have embraced 
approximate variational inference (VI) of the \added{intractable} DGP posterior 
\citep{salimbeni2017doubly}, but this seemingly thrifty approximation 
can sacrifice performance \citep{havasi2018inference,sauer2023vecchia}.
\added{Given our focus on surrogate modeling of expensive deterministic computer
experiments, where data sizes are limited and UQ is essential,} we prefer 
full posterior integration of the latent warping through
Markov chain Monte Carlo (MCMC) sampling.  \added{Previous works have 
found elliptical slice sampling \cite[ESS;][]{murray2010elliptical} to 
consistently outperform VI alternatives on surrogate modeling tasks
\citep{sauer2023vecchia,booth2024non}; we offer a comparison in 
Supplement \ref{supp:existing}.}  

\added{To facilitate gradient-enhancement and gradient predictions,} 
we propose an upgraded deep Gaussian process formulation with gradients on
each Gaussian layer.  We detail how MCMC posterior sampling of the gradient 
of the latent Gaussian layer, combined with careful application of the 
multivariate chain rule, facilitates both gradient-enhancement and posterior 
predictions of the DGP's gradient.  To our knowledge, gradient-enhanced
DGPs have yet to be studied.  Predictions of DGP gradients were
recently considered by \citet{yang2025distribution}, who use strategic
moment approximations to avoid handling gradients on the latent layer.
Our comprehensive Bayesian treatment of the DGP's gradient differs from 
\citeauthor{yang2025distribution}'s thriftier moment-matching approach.

The challenges facing traditional GPs with gradients---namely, ill-conditioning and 
computational bottlenecks---are exacerbated in a DGP.  To address ill-conditioned
covariance matrices, we employ a jitter term \added{on each Gaussian variable};
jitter is a common tool to
preserve numerical stability in deterministic GP regression \citep{gramacy2012cases}.
The simplicity of this solution is \added{preferable} given our complex DGP model, and it 
proved sufficient for all of our test cases.  To alleviate the computational
expense of matrix inversion \added{for larger data sizes} (which \added{are
compounded by} the addition of gradients and GP layers), we implement \added{optional}
Vecchia approximation \citep{vecchia1988estimation}
throughout our upgraded DGP framework.  Vecchia approximation enables faster
GP sampling, likelihood evaluation, and posterior predictions \added{when data sizes
are in the hundreds or more \citep{katzfuss2020vecchia,katzfuss2021general}.  When
gradients are involved, the dimension of covariance matrices may reach the hundreds
even with small data sizes and moderate input dimension (more on this in Section
\ref{sec:review}). Vecchia approximation has 
been shown to outperform alternatives like local or low rank approximations 
with both GP and DGP surrogates 
\citep[e.g.,][]{katzfuss2022scaled,wu2022variational,sauer2023active}.}
As an added bonus, Vecchia approximation also helps
alleviate potential ill-conditioning of gradient-enhanced covariance matrices.

The remainder of this manuscript is organized as follows.  Section~\ref{sec:review}
reviews GPs with gradients.  Section~\ref{sec:method} details our upgraded DGP
and the process for obtaining gradient predictions and incorporating gradient-enhancement.
Section~\ref{sec:implement} describes implementation details, including
integration of Vecchia approximation.  \added{Section~\ref{sec:results} benchmarks 
our gradient-enhanced DGPs and DGP gradient predictions (separately and together)
against relevant GP and DGP competitors on a variety of simulation experiments.
Section~\ref{sec:quantum} deploys our methodology on real-world computer
experiments of quantum mechanics.}  Section~\ref{sec:discuss} concludes
with discussion of limitations and extensions.  All methodology is publicly available
in the {\tt deepgp} R-package \citep{deepgp}.  We also provide code to reproduce
all figures and exercises in a public git 
repository.\footnote{\url{https://bitbucket.org/gramacylab/deepgp-ex/}}

%%%%%%%%%%%%%%%%%%%%%%%%%%%%%%%%%%%%%%%%%%%%%%%%%%%%%%%%%%%%%%%%%%%%%
\section{Gaussian Process Foundations}\label{sec:review}
%%%%%%%%%%%%%%%%%%%%%%%%%%%%%%%%%%%%%%%%%%%%%%%%%%%%%%%%%%%%%%%%%%%%%

For input $\mathbf{x}\in\mathbb{R}^D$, the black-box computer model returns 
$y=f(\mathbf{x})$.  Let $X$ denote the row-combined matrix of 
$\mathbf{x}_i = [x_{i1}, \dots, x_{iD}]$
for $i=1,\dots,n$.  Let $\mathbf{y}$ denote the corresponding response
vector, i.e., $\mathbf{y} = f(X)$.  Throughout, we use lowercase letters
to denote scalars, bold lowercase letters to denote vectors, and uppercase
letters to denote matrices.

A Gaussian process prior assumes that response $\mathbf{y}$ observed
at any finite collection of locations $X$ is a realization of a 
multivariate normal distribution (MVN), e.g., 
$\mathbf{y}\sim\mathcal{N}_n(\boldsymbol\mu, \Sigma)$.  Moving forward, we
fix $\boldsymbol\mu = \mathbf{0}$ after centering responses.
The covariance is parameterized as $\Sigma = \tau^2 (K_{00}(X) + g\mathbb{I}_n)$ with
scale parameter $\tau^2$, kernel function $K_{00}(\cdot)$, and nugget $g$
(where $\mathbb{I}_n$ represents the $n\times n$ identity matrix).
The $00$ subscript is superfluous here, but will be 
essential when we introduce derivatives. 
The nugget term captures random error.  While stochastic computer experiments
are increasingly common \citep{baker2022analyzing}, our focus here is on 
deterministic black-box functions.  We thus fix $g=\varepsilon$ throughout, 
where $\varepsilon$ is fixed at a small value for numerical stability.
\added{The magnitude of $\varepsilon$ should be chosen in light of the
marginal variance of $\mathbf{y}$.  Here, we will adopt the common practice
of pre-scaling responses to have approximately unit variance and fixing
$\varepsilon=1\times 10^{-6}$ (a ratio of 1 to 1 million).}

The ${ij}^\textrm{th}$ element of $K_{00}(X)$ denotes the correlation between 
$y_i$ and $y_j$ for all $i=1,\dots,n$ and $j=1,\dots,n$. 
\added{Our methodological contribution requires a twice differentiable kernel,
so we have implemented the Gaussian/squared exponential and the
Mat\`{e}rn-$5/2$ kernels \citep{stein1999interpolation} in our software.
We define the Gaussian kernel as}
\[
\textrm{Corr}(y_i, y_j) = K_{00}(X)^{ij} = K_{00}(\mathbf{x}_i, \mathbf{x}_j)
= \mathrm{exp}\left(-\sum_{d=1}^D \frac{(x_{id} - x_{jd})^2}{\theta_d}\right),
\] 
where $\boldsymbol\theta = [\theta_1, \dots, \theta_D]$ governs
the ``lengthscale'' in each dimension.  \added{Specifics for the Mat\`{e}rn-$5/2$
kernel are provided in Supplement \ref{supp:matern}.}

Although the details of GP inference are rather textbook 
\citep[e.g.,][]{santner2003design,rasmussen2006gaussian,gramacy2020surrogates}, 
we will dive into them here to set the stage for 
later developments.  Our GP log likelihood function is
\begin{equation}\label{eq:gplogl}
\added{\log\mathcal{L}(\mathbf{y}; X) = -\frac{1}{2}\tau^2 - \frac{1}{2}\log|K_{00}(X) + \varepsilon\mathbb{I}_n|
- \frac{1}{2\tau^2}\mathbf{y}^\top \left(K_{00}(X) + \varepsilon\mathbb{I}_n\right)^{-1} \mathbf{y} + c},
\end{equation}
\added{for some constant $c$.}  This likelihood may be used to infer 
estimates of unknown hyperparameters $\tau^2$ and $\boldsymbol\theta$.
To obtain posterior predictions of $\textbf{\textit{y}} = f(\mathcal{X})$ for an 
$n_p\times D$ matrix of predictive locations $\mathcal{X}$, we start by 
stacking training and testing locations, resulting in the GP prior:
\begin{equation}\label{eq:gpstack}
\begin{bmatrix}
\mathbf{y} \\
\textbf{\textit{y}}
\end{bmatrix} \sim \mathcal{N}_{n+n_p}\left(\mathbf{0}, 
\tau^2 \left(K_\textrm{stack} + \varepsilon\mathbb{I}_{n+n_p}\right)\right)
\quad\textrm{where}\quad
K_\textrm{stack} = \begin{bmatrix}
K_{00}(X) & K_{00}(X, \mathcal{X}) \\
K_{00}(\mathcal{X}, X) & K_{00}(\mathcal{X})
\end{bmatrix},
\end{equation}
and the $ij^\textrm{th}$ element of $K_{00}(\mathcal{X}, X)$ contains the
correlation between the $i^\textrm{th}$ element of $\textbf{\textit{y}}$ and the 
$j^\textrm{th}$ element of $\mathbf{y}$.
Then standard MVN conditioning provides the following posterior,
conditioned on the aforementioned hyperparameters:
\begin{equation}\label{eq:gppred}
\textbf{\textit{y}}\mid \mathbf{y} \sim \mathcal{N}_{n_p}\left(\mu^\star, \Sigma^\star\right)
\quad\textrm{where}\quad
\begin{aligned}
\mu^\star &=  K_{00}(\mathcal{X}, X) \left(K_{00}(X) + \varepsilon\mathbb{I}_n\right)^{-1} \mathbf{y} \\
\Sigma^\star &= \tau^2\left(K_{00}(\mathcal{X}) - K_{00}(\mathcal{X}, X) 
    \left(K_{00}(X) + \varepsilon\mathbb{I}_n\right)^{-1} K_{00}(X, \mathcal{X}) \right).
\end{aligned}
\end{equation}
Since these so called ``kriging equations'' will pop up several times in 
Section~\ref{sec:method}, we will use the following shorthand to indicate 
that the response $\textbf{\textit{y}}$
at predictive locations $\mathcal{X}$ conditioned on $\{X, \mathbf{y}\}$ follows 
the Gaussian distribution of Eq.~(\ref{eq:gppred}):
\begin{equation}\label{eq:gp}
\textbf{\textit{y}}\sim \textrm{GP}\left(\mathcal{X} \mid X, \mathbf{y} \right).
\end{equation}

\begin{figure}[ht!]
\centering
\includegraphics[width=\textwidth,trim=10 10 0 0]{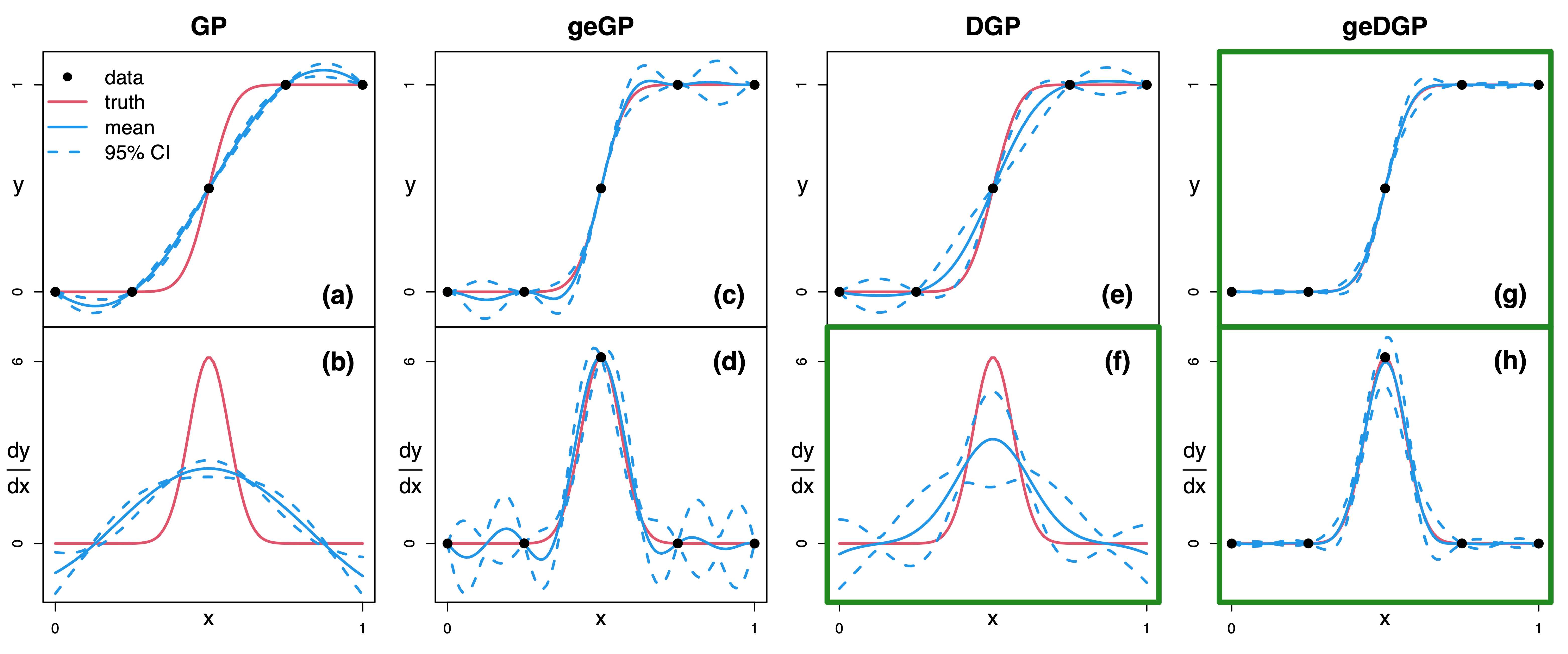}
\caption{Standard and gradient-enhanced GP and DGP predictions
of a simple step function (top) and its gradient (bottom).  The GP
and DGP are only trained on five observations of $y$; the geGP and geDGP
are additionally trained on the corresponding observations of $\frac{dy}{dx}$.
Green boxes highlight the novel contributions of this work.
All models were fit with the {\tt deepgp} package \citep{deepgp}.}
\label{fig:step}
\end{figure}

To demonstrate, consider the function $y = \Phi\left(\frac{x-0.5}{0.065}\right)$ 
with gradient $\frac{dy}{dx} = \frac{1}{0.065}\phi\left(\frac{x-0.5}{0.065}\right)$ 
where $\Phi$ and $\phi$ represent the standard Gaussian CDF and PDF, respectively.  
The red lines in Figure~\ref{fig:step}
portray this ``step'' function (upper panels) and its gradient 
(lower panels) for $x\in[0, 1]$.
Panel (a) shows the GP posterior mean and 95\% credible interval (CI; solid/dashed blue)
conditioned on the five training observations shown in black.  In this case, the
GP provides inaccurate predictions and ineffective UQ.

\subsection{Gradient notation}

Let $\frac{\partial y_i}{\partial x^d}$ denote the partial derivative
of a single observation $y_i$ with respect to dimension $d\in\{1,\dots, D\}$,
and let $\frac{\partial\mathbf{y}}{\partial x^d} = \left[\frac{\partial y_1}{\partial x^d}, \dots, 
\frac{\partial y_n}{\partial x^d} \right]^\top$
denote the $n$-vector of $d^\textrm{th}$ partial derivatives for each observation in $\mathbf{y}$.
Then $\nabla_x \mathbf{y} = \begin{bmatrix}\frac{\partial \mathbf{y}}{\partial x^1} &
\dots & \frac{\partial \mathbf{y}}{\partial x^D}\end{bmatrix}$ of size
$n\times D$ contains the complete gradient information for response vector $\mathbf{y}$.

Let $K_{dj}(\mathbf{x}_i, \mathbf{x}_j)$ denote the correlation
between the $d^\textrm{th}$ partial derivative at location $\mathbf{x}_i$ and
the $j^\textrm{th}$ partial derivative at location $\mathbf{x}_j$. 
We may evaluate the kernel at all possible response/derivative pairings by
differentiating the kernel function, i.e.,
\begin{equation}\label{eq:kernel}
\begin{aligned}
&\mathrm{Corr}\left(\frac{\partial y_i}{\partial x^d}, y_j\right) = 
    K_{d0}(\mathbf{x}_i, \mathbf{x}_j) = 
    K\left(\frac{\partial}{\partial x^d} \mathbf{x}_i, \mathbf{x}_j\right) = 
    \frac{\partial}{\partial x_i^d} K(\mathbf{x}_i, \mathbf{x}_j) \\
&\mathrm{Corr}\left(y_i, \frac{\partial y_j}{\partial x^d}\right) = 
    K_{0d}(\mathbf{x}_i, \mathbf{x}_j) = 
    K\left(\mathbf{x}_i, \frac{\partial}{\partial x^d} \mathbf{x}_j\right) = 
    \frac{\partial}{\partial x_j^d} K(\mathbf{x}_i, \mathbf{x}_j) \\
&\mathrm{Corr}\left(\frac{\partial y_i}{\partial x^d}, \frac{\partial y_j}{\partial x^d}\right) = 
    K_{dd}(\mathbf{x}_i, \mathbf{x}_j) = 
    K\left(\frac{\partial}{\partial x^d} \mathbf{x}_i, \frac{\partial}{\partial x^d} \mathbf{x}_j\right) = 
    \frac{\partial^2}{\partial x_i^d \partial x_j^d} K(\mathbf{x}_i, \mathbf{x}_j) \\
&\mathrm{Corr}\left(\frac{\partial y_i}{\partial x^d}, \frac{\partial y_j}{\partial x^f}\right) = 
    K_{df}(\mathbf{x}_i, \mathbf{x}_j) = 
    K\left(\frac{\partial}{\partial x^d} \mathbf{x}_i, \frac{\partial}{\partial x^f} \mathbf{x}_j\right) = 
    \frac{\partial^2}{\partial x_i^d \partial x_j^f} K(\mathbf{x}_i, \mathbf{x}_j) \quad \textrm{for}\;d \neq f.
\end{aligned}
\end{equation}
\added{Detailed derivations for the Gaussian and Mat\`{e}rn-$5/2$ kernels are 
provided in Supplement \ref{supp:gaussian} and \ref{supp:matern}, respectively.}
Let $K_{df}(X)$ for $d\in\{0,\dots, D\}$ and $f\in\{0,\dots,D\}$ denote the $n\times n$ matrix 
with $ij^\textrm{th}$ element $K_{df}(\mathbf{x}_i, \mathbf{x}_j)$.
Moving forward, we will often work with the response and all partial derivatives simultaneously.
We will use a subscript of ``$\cdot$'' to represent the integers $\{0, 1, \dots, D\}$.

\subsection{GP gradient predictions}

As long as the kernel is twice differentiable, derivatives of a Gaussian process 
are themselves Gaussian processes.
Since the covariance between all possible response-derivative 
pairings is computable (Eq.~\ref{eq:kernel}), 
we may use straightforward applications of ``stacked'' priors 
(Eq.~\ref{eq:gpstack}) and MVN conditioning to obtain posterior 
predictive distributions of the gradient $\nabla_x \textbf{\textit{y}}$ at 
unobserved input locations $\mathcal{X}$.  We prefer to group $\textbf{\textit{y}}$
and all its partial derivatives into a single stacked vector, which has the
following joint posterior:
\begin{equation}\label{eq:gpgradpred}
\textbf{\textit{y}}_\textrm{all} = 
\begin{bmatrix}
\textbf{\textit{y}} \\ \frac{\partial\textbf{\textit{y}}}{\partial x^1} \\ \vdots \\ 
\frac{\partial\textbf{\textit{y}}}{\partial x^D}
\end{bmatrix} \Big|\, \mathbf{y} \sim 
\mathcal{N}_{N_p}\left(\mu^\star, \Sigma^\star\right)
\quad\textrm{where}\quad
\begin{aligned}
\mu^\star &=  K_{\cdot 0}(\mathcal{X}, X) \left(K_{00}(X) + 
\varepsilon\mathbb{I}_n\right)^{-1} \mathbf{y} \\
\Sigma^\star &= \tau^2\left(K_{\cdot \cdot}(\mathcal{X}) - K_{\cdot 0}(\mathcal{X}, X) 
\left(K_{00}(X)+\varepsilon\mathbb{I}_n\right)^{-1} K_{0 \cdot}(X, \mathcal{X}) \right).
\end{aligned}
\end{equation}
Here, $N_p = n_p + n_pD$, and $K_{\cdot 0}(\mathcal{X}, X) = K_{0 \cdot}(X, \mathcal{X})^\top = 
\begin{bmatrix} K_{00}(\mathcal{X}, X) & K_{01}(\mathcal{X}, X) & \dots & 
K_{0D}(\mathcal{X}, X) \end{bmatrix}$.  The form of this posterior distribution 
will also feature throughout Section~\ref{sec:method}; we will refer to it
as simply:
\begin{equation}\label{eq:gpall}
\textbf{\textit{y}}_\textrm{all}\sim \textrm{GP}_\textrm{all}\left(
\mathcal{X} \mid X, \mathbf{y}\right).
\end{equation}
To provide a visual, panel (b) of Figure~\ref{fig:step}
shows the GP's posterior distribution of $\frac{dy}{dx}$ in this one-dimensional setting.
With only 5 observations of $y$, the GP is not able to effectively estimate the
gradient.  Similar to panel (a), the GP is again overconfident in its inaccurate predictions.

\subsection{Gradient-enhanced GPs}

Now presume our black-box function is equipped to return both response and
gradient information, i.e., $\{y, \nabla_x y\} = f(\mathbf{x})$.  This capability
is a common feature of computer simulation experiments, such as those
involving computational fluid dynamics where adjoint solvers can be configured
to return gradient information \citep[e.g.,][]{jacobson2021adjoint,stanford2022gradient}.  
For the same evaluation budget ($n$), 
training data is now upgraded from $\{X, \mathbf{y}\}$ to
$\{X, \mathbf{y}, \nabla_x\mathbf{y}\}$.  We will refer to surrogates trained 
on both response and gradient observations as ``gradient-enhanced.''

Upgrading a GP to additionally condition on gradient observations is possible 
(again as long as the kernel is twice differentiable), but it requires some tedious
notation.  In our framework, a gradient-enhanced GP (geGP) prior may be represented as
\begin{equation}\label{eq:geprior}
\mathbf{y}_\textrm{all} = \begin{bmatrix} \mathbf{y} \\ \frac{\partial\mathbf{y}}{\partial x^1} \\ 
  \vdots \\ \frac{\partial\mathbf{y}}{\partial x^D} \end{bmatrix} \sim \mathcal{N}_N
\left(\mathbf{0}, \;\tau^2\left(K_{\cdot\cdot}(X) + \varepsilon\mathbb{I}_N\right)\right)
\quad\textrm{where}\quad 
K_{\cdot\cdot}(X) = \begin{bmatrix} K_{00}(X) & K_{01}(X) & \dots & K_{0D}(X) \\ 
  K_{10}(X) & K_{11}(X) & \dots & K_{1D}(X) \\
  \vdots & \ddots & &\\
  K_{D0}(X) & K_{D1}(X) & \dots & K_{DD}(X) \end{bmatrix},
\end{equation}
and $N=n+nD$.  Note, we have made a careful and intentional choice to 
include the jitter term along the entire diagonal
of the covariance matrix.  The ill-conditioning of the $K_{\cdot\cdot}(X)$ matrix is
a well-known problem for geGPs.  There have been several workarounds
proposed in the literature, including pivoting \citep{dalbey2013efficient}, 
approximating the covariance with random feature expansions \citep{he2018instability},
and rescaling $\boldsymbol\theta$ \citep{marchildon2023non}.
In our setting, we find the addition of jitter, which is common practice with GP surrogates
\citep{gramacy2012cases}, to be sufficient and more straightforward than the 
aforementioned approaches.

Likelihood-based inference for kernel hyperparameters in a geGP will utilize
\begin{equation}\label{eq:gegplogl}
\added{\log\mathcal{L}\left(\mathbf{y}_\textrm{all}; X\right)
= -\frac{1}{2}\tau^2 - \frac{1}{2}\log|K_{\cdot\cdot}(X) + \varepsilon\mathbb{I}_N| - 
\frac{1}{2\tau^2}\mathbf{y}_\textrm{all}^\top 
\left(K_{\cdot\cdot}(X) + \varepsilon\mathbb{I}_N\right)^{-1}\mathbf{y}_\textrm{all} + c},
\end{equation}
and posterior predictions of $\textbf{\textit{y}} = f(\mathcal{X})$ will follow
\begin{equation}\label{eq:gegppred}
\textbf{\textit{y}} \mid \mathbf{y}_\textrm{all} \sim \mathcal{N}_{n_p}\left(
\mu^\star, \Sigma^\star\right)
    \quad\textrm{where}\quad
\begin{aligned}
\mu^\star &=  K_{0\cdot}(\mathcal{X}, X) 
\left(K_{\cdot\cdot}(X) + \varepsilon\mathbb{I}_N\right)^{-1} \mathbf{y}_\textrm{all} \\
\Sigma^\star &= \tau^2\left(K_{00}(\mathcal{X}) - 
K_{0\cdot}(\mathcal{X}, X) \left(K_{\cdot\cdot}(X) + \varepsilon\mathbb{I}_N\right)^{-1} 
K_{\cdot 0}(X, \mathcal{X}) \right).
\end{aligned}
\end{equation}
The form of this posterior will also feature later on, so we will refer to it as:
\begin{equation}\label{eq:gpge}
\textbf{\textit{y}}\sim \mathrm{GP}_\textrm{ge}\left(\mathcal{X}\mid X, \mathbf{y}_\textrm{all}\right).
\end{equation}
Revisiting Figure~\ref{fig:step}, panel (c) shows geGP predictions.  The gradients convey 
relevant information resulting in a much improved fit.   Nevertheless, the 
GP is still restricted by the stationarity of the kernel and arguably 
overinflates variance in the flat regions to compensate for the 
steep transition in the center.

While we have introduced gradient predictions and gradient-enhancement separately,
they may be easily combined.  To obtain the posterior distribution of 
a geGP's gradients, simply replace 
$K_{\cdot 0}(\mathcal{X}, X)\rightarrow K_{\cdot\cdot}(\mathcal{X}, X)$, 
$K_{00}(X)\rightarrow K_{\cdot \cdot}(X)$, $\mathbf{y}\rightarrow\mathbf{y}_\textrm{all}$,
and $n\rightarrow N$ in Eq.~(\ref{eq:gpgradpred}).  Panel (d) of Figure~\ref{fig:step} 
shows the predictions of the geGP's gradients. Gradient-enhancement massively improves 
the accuracy and UQ of the GP's gradient prediction.

%%%%%%%%%%%%%%%%%%%%%%%%%%%%%%%%%%%%%%%%%%%%%%%%%%%%%%%%%%%%%%%%%%%%%
\section{Deep Gaussian Processes with Gradients}\label{sec:method}
%%%%%%%%%%%%%%%%%%%%%%%%%%%%%%%%%%%%%%%%%%%%%%%%%%%%%%%%%%%%%%%%%%%%%

In this section we provide a comprehensive Bayesian framework that integrates gradient 
information within a DGP.  We start with a brief review of traditional DGPs to provide
context for our novel developments regarding gradients.  Mirroring the structure of 
Section~\ref{sec:review}, we will detail gradient predictions and gradient-enhancement
separately before joining them together.

A deep Gaussian process is simply a functional composition of Gaussian processes.
This definition is inherently broad, enabling varying 
degrees of complexity regarding structure, width, depth, etc. \citep{dunlop2018deep}.
For this work, we will build upon the following DGP prior:
\begin{equation}\label{eq:dgp}
\underbrace{
\mathbf{y} \mid W \sim \mathcal{N}_n\left(\mathbf{0}, 
\tau^2 \left(K_{00}(W) + \varepsilon\mathbb{I}_n\right)\right)
}_\textrm{outer layer}
\quad\quad
\underbrace{
\mathbf{w}_d \stackrel{\textrm{ind}}{\sim} \mathcal{N}_n\left(\mathbf{0}, 
K_{00}(X) + \varepsilon\mathbb{I}_n \right) 
\quad d=1,\dots,D 
}_\textrm{inner layer}
\end{equation}
where $W = \begin{bmatrix} \mathbf{w}_1 &  \mathbf{w}_2 & \dots &  \mathbf{w}_D \end{bmatrix}$.
We will often refer to $\mathbf{w}_d$ as a latent ``node.''
The inner layer serves as a prior distribution over the spatial warping $W$.
The outer layer serves as the likelihood, tying the warping to the observed $\mathbf{y}$.

There are several key assumptions baked into this prior, all of which are common
choices for DGP surrogate modeling \citep{sauer2023deep}.  First, as explained in 
Section~\ref{sec:intro}, we only consider a single inner layer.
Second, we force the dimension of the latent layer $W$ to match the dimension of $X$.
Third, we impose conditional independence among all latent nodes.  
Fourth, we set latent $\mathbf{w}_d$ to be noise free (recall $\varepsilon$ is fixed
at a small value) with unit variance, only including a $\tau^2$ parameter on the outermost GP. 
Fifth, we use a prior mean of zero on 
each $\mathbf{w}_d$ (our software does support setting the prior mean
of $\mathbf{w}_d$ to the $d^\textrm{th}$ column of $X$, but we have found the
zero mean to be appropriate for the nonstationary functions entertained here).
Finally, although not directly represented in the notation above, 
we allow each GP ($D$-many on the inner layer and one on the outer layer) its own 
isotropic lengthscale parameter.  For the remainder of this section, we presume 
kernel hyperparameters ($\tau^2$ and 
all $\theta$'s) are known, directing our focus to latent $W$.  
We will provide further implementation details regarding these hyperparameters 
in Section~\ref{sec:implement}. 

We use elliptical slice sampling \citep{murray2010elliptical} to infer
latent $W$.  Given an initial or previous sample,
$W^\textrm{prev} = \begin{bmatrix} \mathbf{w}_1^{(t-1)} &  \mathbf{w}_2^{(t-1)} & \dots &  
\mathbf{w}_D^{(t-1)} \end{bmatrix}$, ESS proceeds as follows.  For the first node,
draw a random sample $\mathbf{w}_1^\star$ from the prior (the inner Gaussian layer in Eq.~\ref{eq:dgp}).
Then propose $\mathbf{w}_1^{(t)} = \mathbf{w}_1^{(t-1)}\cos(\gamma) + \mathbf{w}_1^\star\sin(\gamma)$
for a randomly selected $\gamma\in[0,2\pi]$.  Accept based on the likelihood
ratio of the outer Gaussian layer given $W^{(t)} = \begin{bmatrix} \mathbf{w}_1^{(t)} &
\mathbf{w}_2^{(t-1)} & \dots &  \mathbf{w}_D^{(t-1)} \end{bmatrix}$ versus $W^\textrm{prev}$.
If rejected, shrink $\gamma$ \citep[following][]{murray2010elliptical}, 
recalculate $\mathbf{w}_1^{(t)}$, and repeat until acceptance is reached.  We iterate
through this entire ESS procedure for each node in a Gibbs framework,
where $W^\textrm{prev}$ contains accepted $\mathbf{w}_i^{(t)}$ for $i < d$ and 
previously sampled $\mathbf{w}_i^{(t-1)}$ for $i\geq d$, and $W^{(t)}$ follows suit
with only the $d^\textrm{th}$ column updated to $\mathbf{w}_d^{(t)}$.
Moving forward, let $t\in\mathcal{T}$ denotes ESS iterations that
have sufficiently burned-in (and optionally been thinned).

\added{ESS offers several advantages which are uniquely suited for the DGP
framework.  Most notably, it enables holistic sampling of each node, as opposed
to algorithms that propose adjustments to smaller chunks
of $\mathbf{w}^{(t)}_d$.  Additionally, it is free of tuning parameters and does not 
require any manual tuning.  It is also rejection-free.  Each iteration will 
result in a new sample thanks to the inner loop where $\gamma$ is shrunk 
until acceptance is reached.  In practice, we find that acceptance often 
occurs with fewer than 10 refinements of $\gamma$.  Finally, if the posterior
happens to be multimodal, the random samples from the prior can facilitate
proposals that effectively ``bounce back and forth'' between modes
\citep{sauer2023active,havasi2018inference}.  As such, the only
user-specified ``setting'' that will significantly affect burn-in is the
initial value $W^{(0)}$.  As an $n\times D$ matrix, this could be
hard to choose, but we find that an identity initialization of $W^{(0)} = X$
is intuitive and works well in practice.  The identity warping
recovers a stationary GP.  We only want to depart from stationarity when
the training data indicates nonstationarity, and in that case we prefer the
simplest warping (i.e., the closest to the identity) that accommodates 
the response surface.}

\begin{figure}[ht!]
\centering
\includegraphics[width=0.95\textwidth, trim=0 0 0 40, clip=True]{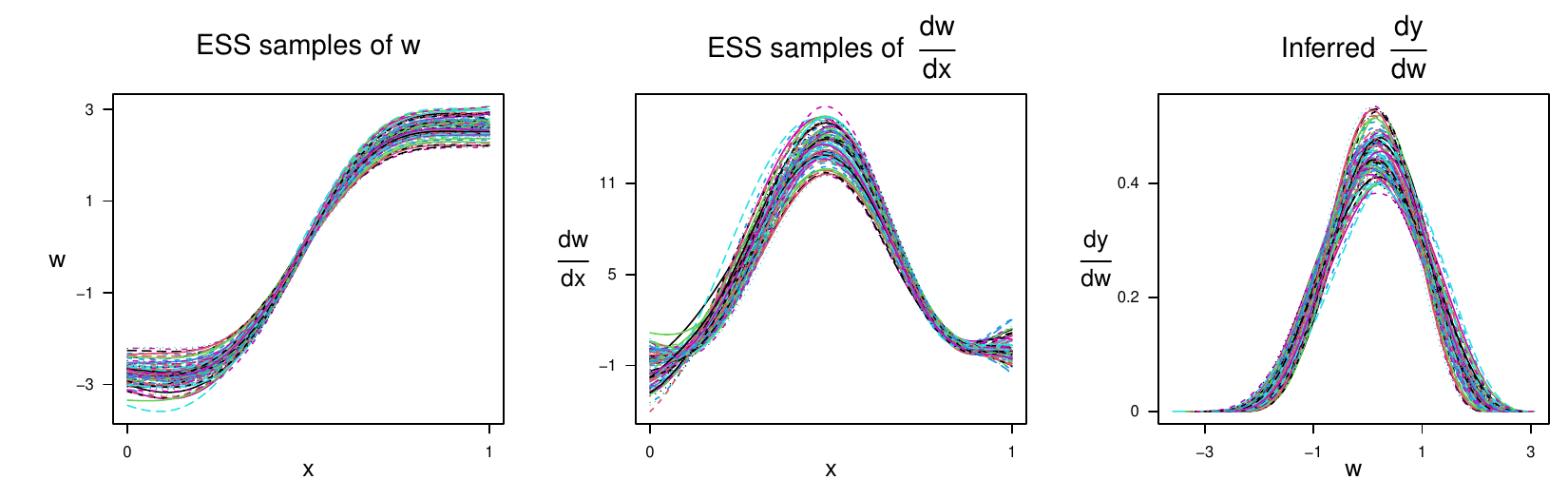}
\caption{{\it Left:} 100 ESS samples of latent $\mathbf{w}$ for a DGP fit to the step function 
from Figure~\ref{fig:step}.  ESS samples have been centered
with some flipped to provide a cleaner visual (in their raw form many samples are
``upside down'' due to the random \added{draw from the prior}, but negating these samples has
no effect on their pairwise distance structure which features in the outer GP layer).
{\it Center:} Corresponding $\frac{d\mathbf{w}}{dx}$.  For unobserved predictive
locations, we infer these (Section~\ref{ss:dgppred}).  For observed training
locations, we sample these (Section~\ref{ss:gedgp}).
{\it Right:} Resulting $\frac{d\mathbf{y}}{dx}$.  For gradient-enhancement 
this is solved according to Eq.~(\ref{eq:solve});
for gradient predictions this is inferred according to Eq.~(\ref{eq:ydeepgradpred}).}
\label{fig:ess}
\end{figure}

To demonstrate, the left panel of Figure~\ref{fig:ess} shows 100 burned-in
posterior samples of $\mathbf{w}$ from a DGP fit to the step function 
from Figure~\ref{fig:step}.
Each ESS sample compresses inputs for
low and high $x$ values where the function is flat and stretches inputs
in the middle of the space where the signal is high.  When viewed over this
warped space (not shown, but resembling an S-curve with a gradual transition instead
of a steep incline), the function from Figure~\ref{fig:step} is relatively stationary.
\added{While ESS samples are easy to visualize when $D=1$, it can be tricky to
assess burn-in in higher dimensions.  For $D>1$, we recommend investigating
trace plots of the outer GP log likelihood (outer layer of Eq.~\ref{eq:dgp}),
which will continue to climb upwards if ESS samples of $W$ are still converging.
See Supplement \ref{supp:trace} for examples.  Ultimately, we seek ESS samples
that optimize predictive performance rather than prioritizing the identification
of an interpretable latent warping, which is an interesting avenue
for future work.}

To predict at new locations, we first infer the warped version of predictive locations 
$\mathcal{X}$ using traditional MVN conditioning (Eq.~\ref{eq:gppred}):
\[
\mathcal{W}^{(t)} = \begin{bmatrix}
\textbf{\textit{w}}_1^{(t)} & \textbf{\textit{w}}_2^{(t)} & \dots & \textbf{\textit{w}}_D^{(t)}
\end{bmatrix}
\quad\textrm{where}\quad
\textbf{\textit{w}}_d^{(t)} \sim \textrm{GP}\left(\mathcal{X} \mid X, \mathbf{w}_d^{(t)}\right)
\quad\textrm{for}\quad
d=1,\dots,D \quad\textrm{and}\quad
t\in\mathcal{T}.
\]
We may draw a joint posterior sample from each of these distributions, but it is also 
common to use just the posterior mean. Our software supports both options.
Then, posterior predictions for $\textbf{\textit{y}}$ follow:
\[
\textbf{\textit{y}}^{(t)} \sim \textrm{GP}\left(\mathcal{W}^{(t)} \mid W^{(t)}, \mathbf{y}, \right)
\quad\textrm{for}\quad
t\in\mathcal{T}.
\]
To aggregate across MCMC iterations, we may accumulate posterior samples for each
$t$, or we may take expectation over $t$ to obtain posterior moments following 
the law of total expectation and variance \citep{sauer2023active}.
Revisiting Figure~\ref{fig:step}, panel (e) shows the DGP fit to the 
simple step function.  Compared to the stationary GP in panel
(a), the DGP provides more accurate predictions with more
effective UQ.  It is a better surrogate for this nonstationary function.

\subsection{Upgrading the model}

Our methodology stems from framing the DGP as a 
multivariate transformation of variables.  The response $\mathbf{y}$ is a function 
of the $D$-dimensional $W$, whose nodes are each functions of the $D$-dimensional 
$X$.  To incorporate gradients in the DGP prior, we first upgrade Eq.~(\ref{eq:dgp}) 
in the spirit of Eq.~(\ref{eq:geprior}), appending derivative observations to 
each response vector:
\begin{equation}\label{eq:gedgp}
\underbrace{
\tilde{\mathbf{y}}_\textrm{all} = 
\begin{bmatrix} 
    \mathbf{y} \\ 
    \frac{\partial \mathbf{y}}{\partial w^1} \\
    \vdots \\ 
    \frac{\partial \mathbf{y}}{\partial w^D}
\end{bmatrix}
\sim\mathcal{N}_N\left(\mathbf{0}, 
\tau^2 \left(K_{\cdot\cdot}(W) + \varepsilon\mathbb{I}_N\right)\right)
}_\textrm{outer layer}
\quad\quad
\underbrace{
\mathbf{w}_{d,\textrm{all}} = 
\begin{bmatrix} 
    \mathbf{w}_d \\ 
    \frac{\partial\mathbf{w}_d}{\partial x^1} \\ 
    \vdots \\ 
    \frac{\partial\mathbf{w}_d}{\partial x^D}
\end{bmatrix}
\sim\mathcal{N}_N\left(\mathbf{0},
K_{\cdot\cdot}(X)+\varepsilon\mathbb{I}_N\right) \quad d=1,\dots, D.
}_\textrm{inner layer} 
\end{equation}
Notice, $\tilde{\mathbf{y}}_\textrm{all}$ which contains $\nabla_w\mathbf{y}$
(the gradient of $\mathbf{y}$ with respect to each node of $W$) differs from 
$\mathbf{y}_\textrm{all}$ as defined in Eq.~(\ref{eq:geprior}) which contains
$\nabla_x\mathbf{y}$.
Each $\mathbf{w}_{d,\textrm{all}}$ contains all the information for a single 
node of the latent warping (the warped values themselves and 
all their partial derivatives with respect to $X$).  To keep notation 
manageable, we will denote the column-binded matrix of these as
\begin{equation}\label{eq:wall}
W_\textrm{all} = 
\begin{bmatrix}
\mathbf{w}_{1,\textrm{all}} & \mathbf{w}_{2,\textrm{all}} & 
    \dots & \mathbf{w}_{D,\textrm{all}}
\end{bmatrix} = 
\begin{bmatrix}
\mathbf{w}_1 & \mathbf{w}_2 & \dots & \mathbf{w}_D \\ 
    \frac{\partial\mathbf{w}_1}{\partial x^1} & 
    \frac{\partial\mathbf{w}_2}{\partial x^1}
    & \dots & \frac{\partial\mathbf{w}_D}{\partial x^1}\\ 
    \vdots \\ 
    \frac{\partial\mathbf{w}_1}{\partial x^D} & 
    \frac{\partial\mathbf{w}_2}{\partial x^D}
    & \dots & \frac{\partial\mathbf{w}_D}{\partial x^D}
\end{bmatrix}.
\end{equation}
Only the first $n$ rows of $W_\textrm{all}$ constitute the $W$ that 
features in $K_{\cdot \cdot}(W)$ in the outer layer of our gradient-DGP.  
The remaining rows make up $\nabla_x W$.

\begin{figure}[ht!]
\centering
\includegraphics[width=0.9\textwidth]{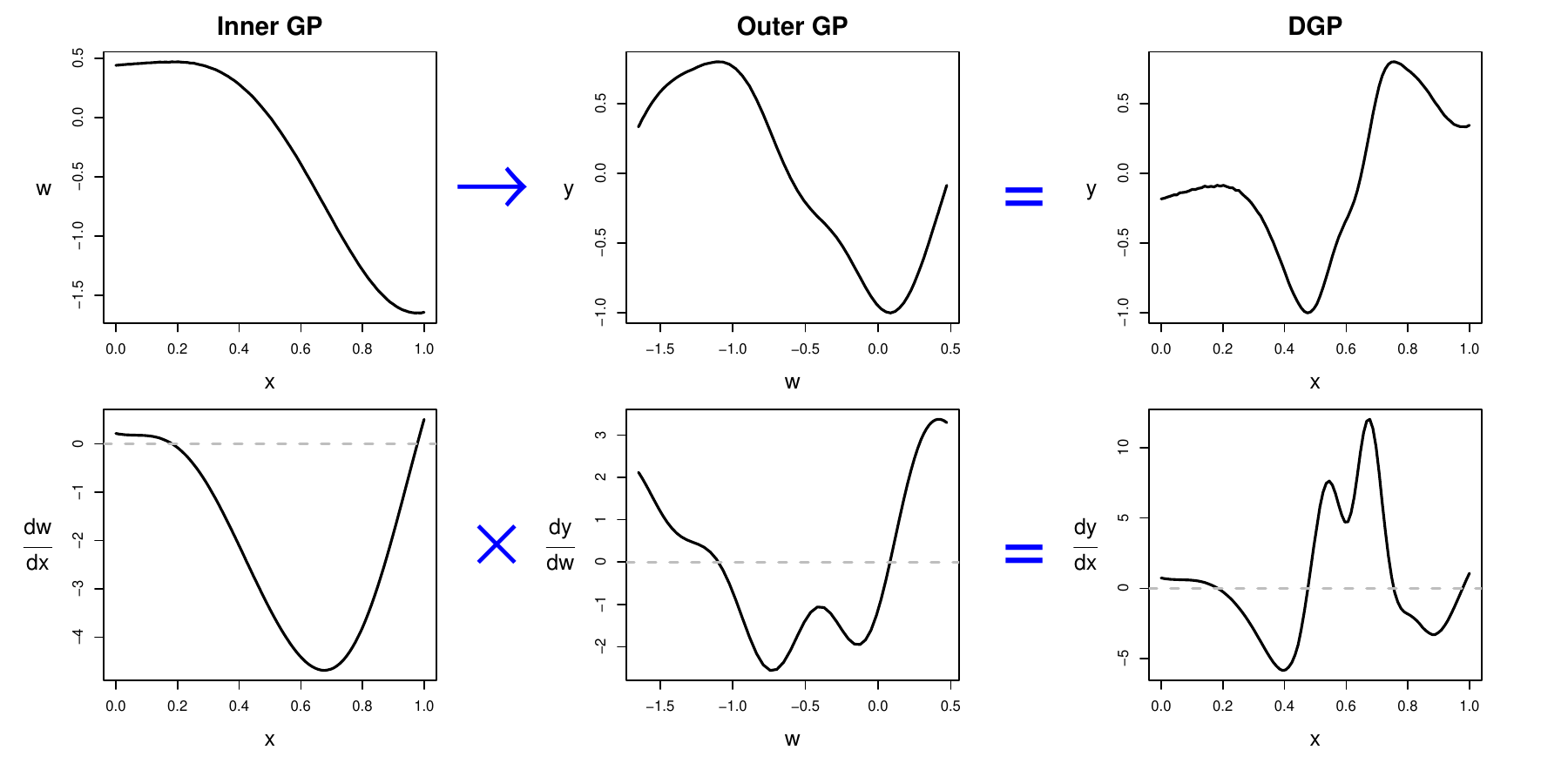}
\caption{A random sample from the gradient-DGP of Eq.~(\ref{eq:gedgp}).
The response and gradient on each layer are stationary GPs;
together they create a nonstationary DGP.}
\label{fig:prior}
\end{figure}

To demonstrate the flexibility and compositional nature of this gradient-DGP
``prior,'' Figure~\ref{fig:prior} shows a simple one-dimensional sample
from the hierarchical model of Eq.~(\ref{eq:gedgp}).  Starting with a 
joint random sample 
of $\mathbf{w}$ and its gradient from a typical GP (left panels), the
sampled $\mathbf{w}$ is fed as input to the another GP, where the response
$\mathbf{y}$ and its gradient are jointly sampled (center panels).  The
response $\mathbf{y}$ viewed over original inputs $\mathbf{x}$ now
comprises a deep Gaussian process (upper right panel).  \added{Its} gradient is 
formed from the product of $\frac{d\mathbf{w}}{dx}$ and
$\frac{d\mathbf{y}}{dw}$.  

The gradient of $\mathbf{y}$ with respect to $X$ ($\nabla_x\mathbf{y}$) is 
notably absent from our gradient-DGP prior.  Rather, it is present through
careful combination of the derivatives on the outer layer ($\nabla_w\mathbf{y}$) 
and the derivatives on the inner layer ($\nabla_x W$).
For a single observation, the multivariate chain rule \citep{corral2013vector} provides
\[
\frac{\partial y}{\partial x^d} = \sum_{i=1}^D 
\frac{\partial y}{\partial w^i} 
\frac{\partial w_i}{\partial x^d}
\quad\textrm{for all}\;\; d=1,\dots,D,
\]
which we condense into the following system of linear equations:
\begin{equation}\label{eq:solve}
\underbrace{
\begin{bmatrix}
\frac{\partial w_1}{\partial x^1} & \frac{\partial w_2}{\partial x^1}
    & \dots & \frac{\partial w_D}{\partial x^1}\\ 
\vdots \\ 
\frac{\partial w_1}{\partial x^D} & \frac{\partial w_2}{\partial x^D}
& \dots & \frac{\partial w_D}{\partial x^D},
\end{bmatrix}
}_{\nabla_x W}
\underbrace{
\begin{bmatrix}
    \frac{\partial y}{\partial w^1} \\
    \vdots \\ 
    \frac{\partial y}{\partial w^D}
\end{bmatrix}
}_{\nabla_w y}
= 
\underbrace{
\begin{bmatrix}
    \frac{\partial y}{\partial x^1} \\
    \vdots \\ 
    \frac{\partial y}{\partial x^D}
\end{bmatrix}
}_{\nabla_x y}.
\end{equation}
After inferring $\nabla_x W$ (more on this momentarily), this linear system enables 
us to convert predictions of $\nabla_w\mathbf{y}$ to predictions of $\nabla_x\mathbf{y}$
(Section~\ref{ss:dgppred}) and/or solve for $\nabla_w\mathbf{y}$ when
$\nabla_x\mathbf{y}$ is observed (Section~\ref{ss:gedgp}).

\subsection{DGP gradient predictions}\label{ss:dgppred}

Suppose we observe $\mathcal{D}_n = \{X, \mathbf{y}\}$ and
have conducted traditional DGP ESS sampling to obtain $W^{(t)}$ for $t\in\mathcal{T}$.
With the ``training'' process completed, we may obtain DGP predictions of 
$\textbf{\textit{y}}$ and $\nabla_x\textbf{\textit{y}}$ at predictive locations 
$\mathcal{X}$ through careful application of standard GP gradient predictions
(Eq.~\ref{eq:gpgradpred}).  First, for each iteration and each
node, we infer the warping $\mathcal{X}\rightarrow\mathcal{W}$ 
and its gradient $\nabla_x \mathcal{W}$ through standard MVN conditioning:
\begin{equation}\label{eq:wgradpred}
\textbf{\textit{w}}_{d,\textrm{all}}^{(t)} =
\begin{bmatrix}
\textbf{\textit{w}}_d^{(t)} \\
\frac{\partial \textbf{\textit{w}}_d^{(t)}}{\partial x^1} \\ 
\vdots \\
\frac{\partial \textbf{\textit{w}}_d^{(t)}}{\partial x^D}
\end{bmatrix} \sim 
\textrm{GP}_\textrm{all}\left(\mathcal{X} \mid X, \mathbf{w}_d^{(t)} \right)
\quad\textrm{for}\quad
d=1,\dots,D \quad\textrm{and}\quad
t\in\mathcal{T}.
\end{equation}
Recall, $\mathbf{w}_d^{(t)}$ represents the $d^\textrm{th}$ column of 
sampled $W^{(t)}$.  Next, we 
column-bind across $d=1,\dots,D$ to obtain $\mathcal{W}_\textrm{all}^{(t)}$.
The first $n_p$ rows constitute $\mathcal{W}^{(t)} = \begin{bmatrix}
\textbf{\textit{w}}_1^{(t)} & \dots & \textbf{\textit{w}}_D^{(t)}
\end{bmatrix}$.  The remaining rows make up $\nabla_x \mathcal{W}^{(t)}$.
Then, $\mathcal{W}^{(t)}$ feeds into another GP to infer 
$\textbf{\textit{y}}$ and its gradient
over the warped space, $\nabla_w\textbf{\textit{y}}$:
\begin{equation}\label{eq:ydeepgradpred}
\tilde{\textbf{\textit{y}}}_\textrm{all}^{(t)} = 
\begin{bmatrix}
\textbf{\textit{y}}^{(t)} \\
\frac{\partial\textbf{\textit{y}}^{(t)}}{\partial w^1} \\ 
    \vdots \\ 
    \frac{\partial\textbf{\textit{y}}^{(t)}}{\partial w^D} 
\end{bmatrix}
\sim \textrm{GP}_\textrm{all}
\left(\mathcal{W}^{(t)} \mid W^{(t)}, \mathbf{y}\right)
\quad\textrm{for}\quad
t\in\mathcal{T}.
\end{equation}
Again, we use the tilde symbol to indicate that the partial
derivatives here are with respect to $W$, not $X$.  All 
elements of $\tilde{\textbf{\textit{y}}}_\textrm{all}^{(t)}$ are
indexed by iteration $t$, since they depend on $\mathcal{W}^{(t)}$ and $W^{(t)}$.

If we draw samples directly from this posterior, we may
convert samples of $\nabla_w \textbf{\textit{y}}^{(t)}$ to 
$\nabla_x \textbf{\textit{y}}^{(t)}$ using 
$\nabla_x \mathcal{W}^{(t)}$, which we inferred in Eq.~(\ref{eq:wgradpred}).
For each predictive location $y\in\textbf{\textit{y}}$, we simply compute
$\nabla_x y^{(t)}$ according to Eq.~(\ref{eq:solve}).
Sometimes, instead of working with a full set of posterior samples,
it is helpful to condense things into the first and second posterior 
moments.  To do this, we consider $\mathcal{W}_\textrm{all}^{(t)}$ as
a constant once it has been inferred.  This process of inferring the 
values of the inner layer then treating the outer layer as conditionally
independent is sometimes referred to as ``stochastic imputation'' 
\citep{ming2023deep}.  

Conditioned on elements of $\mathcal{W}_\textrm{all}^{(t)}$ (e.g., 
$\frac{\partial w_i^{(t)}}{\partial x^d}$ for $i=1,\dots,n_p$ and $d=1,\dots,D$),
the expectation and variance of the $d^\textrm{th}$ partial derivative for any 
$y\in{\textbf{\textit y}}$ at iteration $t$ are
\[
\mathbb{E}\left[\frac{\partial y^{(t)}}{\partial x^d}\right] = 
\sum_{i=1}^D \frac{\partial w_i^{(t)}}{\partial x^d} \;
\mathbb{E}\left[\frac{\partial y^{(t)}}{\partial w^i}\right]
\quad\textrm{and}\quad
\mathbb{V}\left[\frac{\partial y^{(t)}}{\partial x^d}\right] = 
\sum_{i=1}^D \left(\frac{\partial w_i^{(t)}}{\partial x^d}\right)^2 
\;\mathbb{V}\left[\frac{\partial y^{(t)}}{\partial w^i}\right]
\]
where the expectations and variances on the right-hand sides follow the form of 
$\mu^\star$ and $\Sigma^\star$ from Eq.~(\ref{eq:gpgradpred})
with $\mathcal{W}^{(t)}$ in place of $\mathcal{X}$ and $W^{(t)}$ in place of $X$.
Ultimately, we may take the expectation over $t\in\mathcal{T}$ to obtain
the summarized moments:
\[
\mathbb{E}\left[\frac{\partial y}{\partial x^d}\right]
= \frac{1}{\mid\mathcal{T}\mid}\sum_{t\in\mathcal{T}} 
\mathbb{E}\left[\frac{\partial y^{(t)}}{\partial x^d}\right]
\quad\textrm{and}\quad
\mathbb{V}\left[\frac{\partial y}{\partial x^d}\right]
= \frac{1}{\mid\mathcal{T}\mid}\sum_{t\in\mathcal{T}} 
\mathbb{V}\left[\frac{\partial y^{(t)}}{\partial x^d}\right] + 
\mathbb{C}\textrm{ov}_t\left(\mathbb{E}\left[
\frac{\partial y^{(t)}}{\partial x^d}\right]\right).
\]
As a simple demonstration, Figure~\ref{fig:ess} shows the gradients on each
layer of the DGP fit to our one-dimensional step function.
The center panel features $\frac{dw}{dx}$, which can be inferred
(Eq.~\ref{eq:wgradpred}) given the ESS samples of the left panel.
The right panel shows $\frac{dy}{dw}$ which can also be inferred
(Eq.~\ref{eq:ydeepgradpred}) given the ESS samples of the left panel.
A simple multiplication of these derivatives provides the 
prediction of $\frac{dy}{dx}$ from our DGP, as shown in panel (f)
of Figure~\ref{fig:step}.  The DGP's gradient prediction is far
better than the GP's gradient prediction for this nonstationary example.

\subsection{Gradient-enhanced DGPs}\label{ss:gedgp}

Now suppose we observe $\{X,\mathbf{y}, \nabla_x\mathbf{y}\}$, and we seek
gradient-enhanced DGP predictions of $\textbf{\textit y}$.  Incorporating
gradients in our DGP training requires modification of our ESS scheme.
The gradient-enhanced version of the likelihood of the outer layer, 
which will be used in our ESS acceptance probability, is
\begin{equation}\label{eq:gedgpll}
\added{\log\mathcal{L}\left(\mathbf{y}_\textrm{all}; W_\textrm{all}\right)
= -\frac{1}{2}\tau^2 - \frac{1}{2}\log|K_{\cdot\cdot}(W) + \varepsilon\mathbb{I}_N| - 
\frac{1}{2\tau^2}\tilde{\mathbf{y}}_\textrm{all}^\top 
\left(K_{\cdot\cdot}(W) + \varepsilon\mathbb{I}_N\right)^{-1}\tilde{\mathbf{y}}_\textrm{all} + c},
\end{equation}
\added{where $\tilde{\mathbf{y}}_\textrm{all}$, which includes the gradient with respect
to $W$, is calculated from the $\nabla_x W$ elements of $W_\textrm{all}$ and
the $\nabla_x\mathbf{y}$ elements of observed $\mathbf{y}_\textrm{all}$.
To obtain $\nabla_x W$, we propose expanding our ESS
procedure to sample $W_\textrm{all}$ (Eq.~\ref{eq:wall}) instead of just $W$.}

One iteration of our gradient-enhanced ESS scheme proceeds as follows.
Starting with an initial or previous sample $W_\textrm{all}^\textrm{prev} = 
\begin{bmatrix} \mathbf{w}_{1,\textrm{all}}^{(t-1)} & \mathbf{w}_{2,\textrm{all}}^{(t-1)}
& \dots & \mathbf{w}_{D,\textrm{all}}^{(t-1)}\end{bmatrix}$, we 
use our observed $\nabla_x \mathbf{y}$ and the $\nabla_x W^{(t-1)}$ 
components of $W_\textrm{all}^\textrm{prev}$ (removing the first $n$ rows) to solve
the system of linear equations in Eq.~(\ref{eq:solve}) for $\nabla_w\mathbf{y}^\textrm{prev}$.
Appending this to the observed $\mathbf{y}$ provides 
$\tilde{\mathbf{y}}_\textrm{all}^\textrm{prev}$ which will be used in our likelihood
evaluations.  Then, starting with the first node, we sample 
$\mathbf{w}_{1,\textrm{all}}^\star$
from the MVN distribution presented in the inner layer of Eq.~(\ref{eq:gedgp}).
Since this prior distribution is still Gaussian, it is suitable for ESS.
Next, we propose as we normally would, but we include the gradients in each 
component:
\begin{equation}\label{eq:proposal}
\mathbf{w}_{d,\textrm{all}}^{(t)} = 
    \mathbf{w}_{d,\textrm{all}}^{(t-1)}\cos(\gamma)
    + \mathbf{w}_{d,\textrm{all}}^\star\sin(\gamma).
\end{equation}
We create $W_\textrm{all}^{(t)}$ by duplicating $W_\textrm{all}^\textrm{prev}$ 
and overwriting the $d^\textrm{th}$ column with 
$\mathbf{w}_{d,\textrm{all}}^{(t)}$.
Then, we again solve the system of linear equations with the updated gradients
from $W_\textrm{all}^{(t)}$ to obtain $\nabla_w\mathbf{y}^{(t)}$ and
$\tilde{\mathbf{y}}^{(t)}$.  With all of these elements in hand, we accept
based on the ratio of the likelihood from Eq.~(\ref{eq:gedgpll})
with $W^{(t)}$ (the first $n$ rows of $W_\textrm{all}^{(t)}$) and 
$\tilde{\mathbf{y}}_\textrm{all}^{(t)}$ versus $W^\textrm{prev}$ and 
$\tilde{\mathbf{y}}_\textrm{all}^\textrm{prev}$.
The rest of the algorithm proceeds as normal, shrinking $\gamma$ until 
acceptance is reached, and iterating through the nodes in a Gibbs fashion.

At the end of our ESS procedure, we will have burned-in posterior samples
$W_\textrm{all}^{(t)}$ and the resulting $\tilde{\mathbf{y}}_\textrm{all}^{(t)}$ 
for $t\in\mathcal{T}$.  To predict $\textbf{\textit{y}} = f(\mathcal{X})$ with
our gradient-enhanced DGP, we again start by finding the appropriate 
warping of the predictive locations.
We use the GP kriging equations on each node, but this time we condition
on $\mathbf{w}_{d,\textrm{all}}^{(t)}$ instead of just $\mathbf{w}_d^{(t)}$:
\begin{equation}\label{eq:wgedgp}
\textbf{\textit{w}}_d^{(t)} \sim \textrm{GP}_\textrm{ge}\left(\mathcal{X} \mid X,
\mathbf{w}_{d,\textrm{all}}^{(t)}\right)
\quad\textrm{for}\quad
t\in\mathcal{T}.
\end{equation}
Note, we do not need to predict gradients here if we only seek predictions
of $\textbf{\textit{y}}$.  We then aggregate these warpings into $\mathcal{W}^{(t)}$
and infer $\textbf{\textit{y}}$ in standard fashion, using $\tilde{\mathbf{y}}_\textrm{all}^{(t)}$,
which we solved for in our ESS procedure:
\begin{equation}\label{eq:ygedgp}
\textbf{\textit{y}}^{(t)}\sim\mathrm{GP}_\textrm{ge}\left(\mathcal{W}^{(t)} \mid 
W^{(t)}, \tilde{\mathbf{y}}_\textrm{all}^{(t)}\right)
\quad\textrm{for}\quad
t\in\mathcal{T}.
\end{equation}
As always, we may draw samples from this posterior or work with the summarized
posterior moments, aggregated across $t\in\mathcal{T}$.

\begin{figure}[ht!]
\centering
\includegraphics[width=\textwidth, trim=0 30 0 0, clip=True]{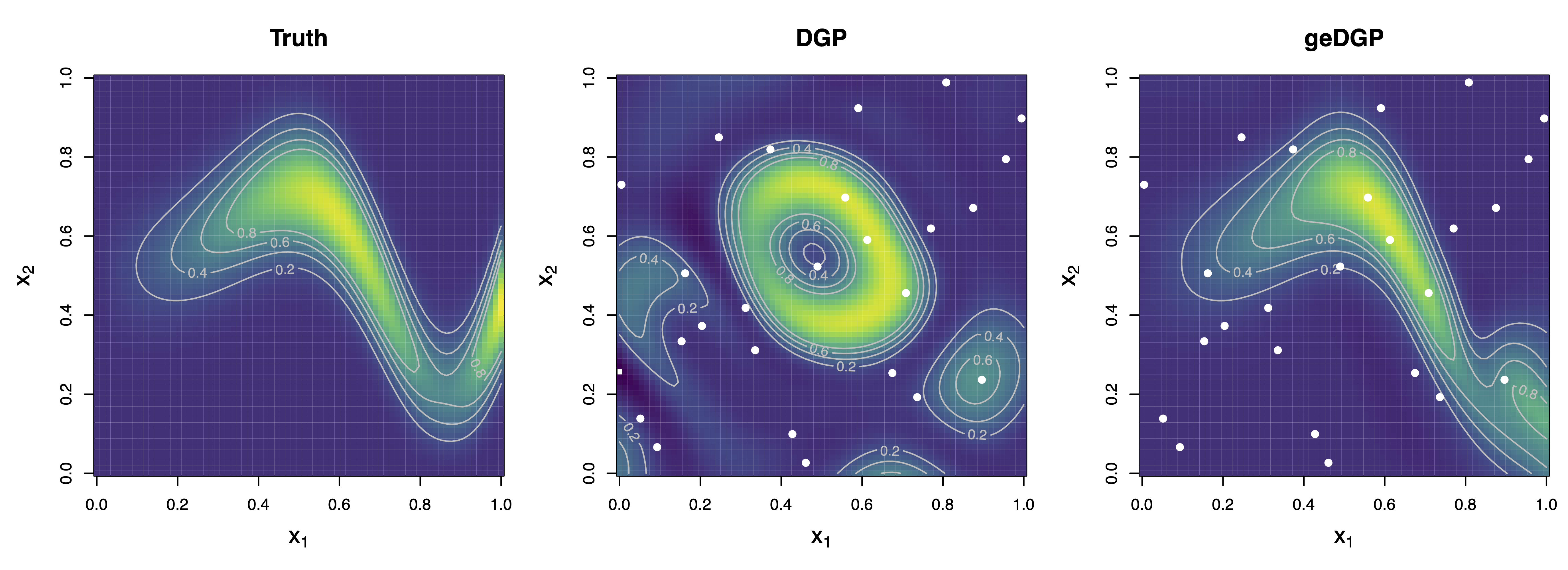}
\caption{Heatmap of the 2d squiggle function (left) from \citet{duqling} with the predicted
mean from a DGP (center) and geDGP (right) trained on the same random sample
of size $n=25$ (white circles).}
\label{fig:squiggle}
\end{figure}

Returning to Figure~\ref{fig:step}, panel (g) shows our gradient-enhanced DGP 
predictions.  The flexibility of the DGP combined with the incorporation of 
gradient observations provides a nearly perfect fit, improving upon both the
standard DGP and the gradient-enhanced GP.  As one more visual of the power
of gradient-enhancement, consider the two-dimensional ``squiggle'' function
\citep{duqling} shown in the left panel of Figure~\ref{fig:squiggle}.  
The center and right panels show the predicted mean from a DGP and geDGP, 
respectively, trained on the same random sample with $n=25$ (white circles).  
Gradient-enhancement hugely improves the accuracy of the surrogate's predictions.
We will feature this function in our simulation exercises of Section~\ref{sec:results},
where we will also show the geDGP consistently provides more effective UQ.

In our framework, it is relatively straightforward to combine our gradient-enhanced
DGP methodology with the procedures outlined in Section~\ref{ss:dgppred} to obtain
gradient-enhanced DGP gradient predictions.  When we infer the warpings
of Eq.~(\ref{eq:wgedgp}), we simply infer the entire 
$\textbf{\textit{w}}_{d,\textrm{all}}^{(t)}$ in place of $\textbf{\textit{w}}_d^{(t)}$.
Then we similarly upgrade Eq.~(\ref{eq:ygedgp}) to infer 
$\tilde{\textbf{\textit{y}}}^{(t)}_\textrm{all}$ in place of $\textbf{\textit{y}}^{(t)}$.
Finally, we combine the inferred gradients of these components ($\nabla_x\mathcal{W}^{(t)}$
and $\nabla_w\textbf{\textit{y}}^{(t)}$) following Eq.~(\ref{eq:solve}) to obtain
$\nabla_x\textbf{\textit{y}}^{(t)}$.  Panel (h) of Figure~\ref{fig:step} shows
the prediction of the gradient from the geDGP fit to the step function.  Again, 
it is a stark improvement from the predictions of the geGP and the DGP.

%%%%%%%%%%%%%%%%%%%%%%%%%%%%%%%%%%%%%%%%%%%%%%%%%%%%%%%%%%%%%%%%%%%%%
\section{Implementation}\label{sec:implement}
%%%%%%%%%%%%%%%%%%%%%%%%%%%%%%%%%%%%%%%%%%%%%%%%%%%%%%%%%%%%%%%%%%%%%

In this section we provide additional implementation details.  All of our methodology
is implemented and freely available in the {\tt deepgp} R-package on CRAN \citep{deepgp}.

\subsection{\added{Computation}}

\added{The added flexibility of a DGP over a stationary GP comes at the cost
of additional computation, largely thanks to the inference of latent $W$.  It is no
surprise that the incorporation of gradients brings with it even higher compuational demands.
In general, the computational cost of training a Bayesian DGP as described in 
Section \ref{sec:method} (i.e., collecting burned-in posterior samples of 
each latent quantity) will increase with 
$n$, $D$, and the number of MCMC iterations.  Each likelihood evaluation is 
computationally cubic in $n$, and the number of likelihood evaluations required scales 
almost linearly with $D$ and the number of MCMC iterations (there is some variability
thanks to the number of refinements of $\gamma$ required to reach acceptance
in each ESS iteration, which will vary for each $t$).  The random draw from the inner
GP used within each ESS proposal also experiences computational costs cubic in $n$.
Gradient-enhancement takes all of these computations from $\mathcal{O}(n^3)$
to $\mathcal{O}(N^3)$ where $N=n+nD$.  Gradient-enhancement also upgrades the matrix
inverses required for predicting at unobserved locations from $n\times n$ to $N\times N$.
Moreover, predicting gradients (with or without gradient-enhancement) adds another $n_p D$ many
quantities that must be inferred.  While we may opt to predict each of these independently 
in parallel, occasionally joint posterior samples are necessary
\citep[say, for example, using Thompson sampling for Bayesian optimization,][]{thompson1933likelihood}.
Joint posterior samples that include gradients can have $\mathcal{O}(N_p^3)$ 
computational costs.}

\added{For surrogate modeling of expensive computer experiments, the greatest computational
burden lies in the evaluation of the computer model, so the cost of training and using
the surrogate is not a major concern.  Nevertheless, strategies to avoid computational
bottlenecks when $N$ or $N_p$ become too large are warranted.  The most effective
strategy we've found is Vecchia approximation, which we will discuss next in Section
\ref{sec:vecchia}, but other simple strategies can still be very useful, such as
using an optmized BLAS/LAPACK \citep{santillan2018improving} and employing parallel
computation (particularly for predicting across $t\in\mathcal{T}$).  If computation
is really a concern, it is also possible to use gradient-enhancement for training
but not for prediction, or vice versa, although this is ill-advised as it could 
impair predictive performance.}

\subsection{Vecchia approximation}\label{sec:vecchia}

\added{In this section we detail our implementation of Vecchia approximation for
DGPs with gradients, which we have integrated as an option across every
facet of our software.  When data sizes are large (typically above several hundred), 
Vecchia approximation can significantly reduce the time required for GP
likelihood evaluation and GP predictions by avoiding computations that are cubic
in $n$, $N$, $n_p$, or $N_p$.  These speed-ups will accumulate in
our Bayesian DGP computations \citep{sauer2023vecchia}.}

The Vecchia approximation \citep{vecchia1988estimation} relies on the fact 
that any joint distribution may be factored into a product of univariate 
distributions, e.g.,
\[
\log\mathcal{L}\left(\mathbf{y} = [y_1, y_2, \dots, y_n]\right) = 
    \log\mathcal{L}(y_1) + 
    \sum_{i=2}^n \log\mathcal{L}(y_i \mid \mathbf{y}_{c_i})
    \quad\textrm{where}\quad
    c_i = \{1, \dots, i-1\}.
\]
When $n$ is large, the Vecchia approximation
takes $c_i \subseteq \{1,\dots,i-1\}$, where the maximum size of $c_i$ is 
capped at some value $m \ll n$.  The approximation is determined by the
ordering of the observations ($y_i$ can only condition on $y_j$ when $j<i$)
and the choice of conditioning sets $c_i$.  Vecchia approximation has been
thoroughly explored for stationary Gaussian processes 
\citep[e.g.,][]{datta2016hierarchical,katzfuss2020vecchia,katzfuss2021general,kang2024asymptotic},
and recently extended to deep Gaussian processes \citep{sauer2023vecchia}.

Upgrading existing methodologies to incorporate gradients required two
modifications: (i) updating under-the-hood covariance calculations to use the
appropriate gradient kernels from Eq.~(\ref{eq:kernel}), and (ii) determining 
an effective ordering and conditioning structure for responses and partial
derivatives together.  The first of these is rather straightforward as long as
we carefully track derivative indices.  We work with the Cholesky 
decomposition of the inverse of each GP covariance matrix
following \citet{katzfuss2021general}, and simply upgrade its calculation
\citep[][Eq.~9]{sauer2023vecchia} to use the appropriate kernel.

The second modification requires a bit more development.  Typical Vecchia
frameworks commonly employ random orderings with conditioning sets based on 
nearest neighbors, although more complicated versions have been entertained
\citep[e.g.,][]{guinness2018permutation,katzfuss2022scaled}.  For our
setting, we propose the following, having found it to work well in our
benchmark exercises.  We defer a thorough exploration of the costs and benefits
of various Vecchia orderings and conditioning sets for
GPs/DGPs with gradients to future work.  Instead of a random ordering of all
responses and gradients together, we require that all response observations be
ordered first.  We randomly order $y_i$ for $i=1,\dots,n$, then append each
$\frac{\partial y_i}{\partial x^d}$ using the same ordering, in turn.
We then select conditioning sets through nearest neighbors with one 
caveat: when there are ties, an observation of $y$ should be 
selected over any gradient observations.  Through the use of these 
smaller conditioning sets, Vecchia approximation offers additional 
protection against potentially ill-conditioned covariance 
matrices in gradient-enhanced models.

\begin{figure}[ht!]
\centering
\includegraphics[width=0.8\textwidth]{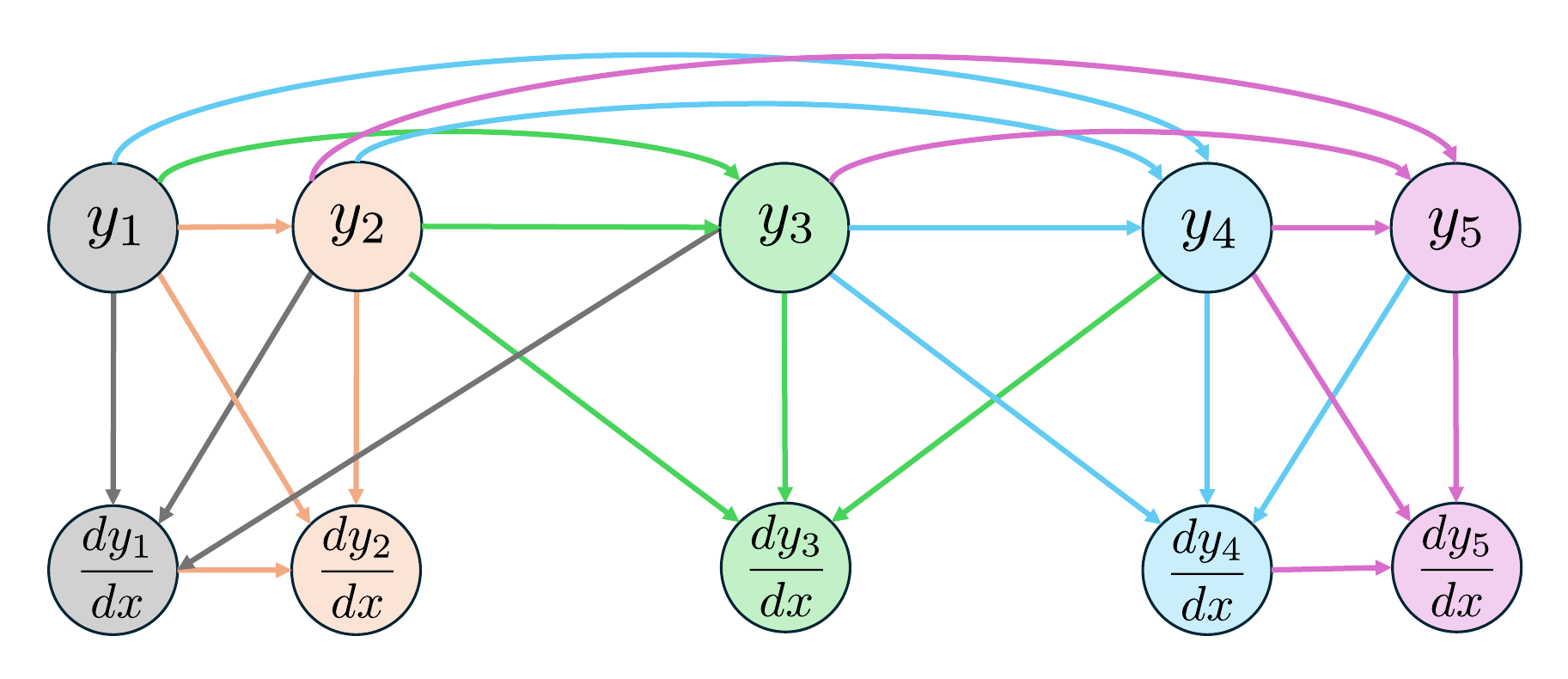} \\
\caption{Vecchia conditioning sets with $n=5$, $D=1$, and $m=3$.}
\label{fig:vecchia}
\end{figure}

Figure~\ref{fig:vecchia} offers a simple example, with ordering
$\left\{y_1, y_2, y_3, y_4, y_5, \frac{dy_1}{dx}, \frac{dy_2}{dx}, \frac{dy_3}{dx},
\frac{dy_4}{dx}, \frac{dy_5}{dx}\right\}$, where the initial allocation of 
indices $1,\dots,5$ was random.  
The conditioning structure is indicated
by the colored arrows (the horizontal spacing is intended to convey distance,
such that $y_2$ is closer to $y_1$ than it is to $y_3$).  Each 
$y_1,\dots y_5$ follows standard nearest neighbors conditioning, selecting 
the closest $m$ observations that were ordered previously.  $\frac{dy_1}{dx}$ similarly
conditions on the nearest $y$ observations.  $\frac{dy_2}{dx}$ is
the first to condition on a derivative, conditioning on $\frac{dy_1}{dx}$
instead of $y_3$ as it is a nearer neighbor.  $\frac{dy_3}{dx}$ obviously
conditions on $y_3$, but it then faces a three-way tie among 
$y_2$, $y_4$, and $\frac{dy_2}{dx}$.  When ties occur,
we prioritize response observations, thereby selecting $y_2$ and $y_4$.  
Conditioning for $\frac{dy_4}{dx}$ and $\frac{dy_5}{dx}$ follows suit.
We have implemented these same ordering and conditioning
procedures for both gradient-enhancement and gradient predictions in our software.

\subsection{Kernel hyperparameters}\label{ss:hyper}

We choose to infer kernel hyperparameters in our gradient-DGP in a 
fully-Bayesian framework to provide thorough uncertainty quantificaton,
but we acknowledge that alternative treatments
\citep[such as expectation maximization,][]{ming2023deep} may work
equally well.  For $\tau^2$ on the outer layer (Eq.~\ref{eq:gedgp}), we 
use a reference prior $\pi(\tau^2)\propto\frac{1}{\tau^2}$, which enables
closed-form integration of $\tau^2$ from the outer Gaussian likelihood
\citep[][Chapter 5]{gramacy2020surrogates}.  For lengthscales (one on the
outer layer inside $K_{\cdot\cdot}(W)$ and $D$-many on the inner layer
inside each $K_{\cdot\cdot}(X)$), we conduct
Metropolis-Hastings (MH) sampling using the Gamma priors and sliding
window proposals outlined in \citet[][Section~5.1]{sauer2023active}.
We integrate MH sampling of lengthscales with our ESS sampling of
latent nodes in one large Gibbs sampling loop.

%%%%%%%%%%%%%%%%%%%%%%%%%%%%%%%%%%%%%%%%%%%%%%%%%%%%%%%%%%%%%%%%%%%%%
\section{Benchmarking}\label{sec:results}
%%%%%%%%%%%%%%%%%%%%%%%%%%%%%%%%%%%%%%%%%%%%%%%%%%%%%%%%%%%%%%%%%%%%%

\added{To validate our methods, we benchmark our contributions 
against existing DGP and gradient-enhanced GP methodologies on a variety
of nonstationary test functions.  At this time, there is no 
alternative software that facilitates gradient-enhancement or gradient
predictions with DGPs, so our focus is on comparing our methodology
to existing non-enhanced DGP and gradient-enhanced GP implementations.
For each function, we conduct 
30 Monte Carlo repetitions with re-randomized Latin hypercube 
sampling \citep[LHS;][]{mckay2000comparison} 
training designs.
We measure performance on a testing LHS of size $100D$. Formulaic details of 
each test function are provided in Supplement \ref{supp:functions}.  
Reproducible code for all exercises is available in our public git 
repository.}\footnote{\url{https://bitbucket.org/gramacylab/deepgp-ex/}}

\added{For existing DGP competitors, we originally considered the fully-Bayesian
DGP of \citet{sauer2023active} implemented in the {\tt deepgp} package 
\citep{deepgp} with both the Gaussian and Mat\`{e}rn-$5/2$ kernels 
and the doubly stochastic variational inference DGP of 
\citet{salimbeni2017doubly} implemented in the {\tt gpflux} python
package \citep{gpflux}.  For existing geGP competitors, we originally
considered the Bayesian implementation offered in the {\tt deepgp}
package (which follows the methodology described in this work),
the gradient-enhanced kriging implementation provided in the python
{\tt Surrogate Modeling Toolbox} \citep{smt}, and the gradient-enhanced
kriging implementation of \citet{hung2021random}.  Further details,
results, and discussion of each of these existing methodologies
are provided in Supplement \ref{supp:existing}.  We found that the Bayesian
DGPs that use ESS universally outperformed the variational inference DGPs,
regardless of the kernel used.  Similarly, the Bayesian geGP from the
{\tt deepgp} package offered consistently better performance than the
alternate gradient-enhanced kriging implementations.  Consequently,
we retain only these best performers as comparators here.}

\added{The following results thus contain four unique surrogates, all 
leveraging Bayesian implementations from the {\tt deepgp} package: 
a stationary GP as a baseline, a gradient-enhanced GP, a nonstationary DGP,
and a nonstationary gradient-enhanced DGP.  We use Gaussian kernels
for each, although the software also offers the Mat\`{e}rn-$5/2$ kernel.  
Since all of these use
the same software with consistent implementation choices, we are able
to isolate the effects of the GP vs.~the DGP and gradient-enhancement 
vs.~non-enhancement.}

\added{For this large simulation study, with 30 repetitions of each model
across 3 functions, we find it infeasible to manually
inspect trace plots to assess burn-in for each Bayesian surrogate.  Instead,
we settle upon default values that proved sufficient in initial testing.
For our GP and geGP surrogates, we conduct 5,000 MCMC iterations, removing 3,000 for
burn-in, then thinning by 2.  For DGP surrogates, which have much more
to infer, we conduct 10,000 MCMC iterations, removing 8,000 for 
burn-in, then thinning by 2.}  We initialize the latent layer of
the DGP and geDGP at the identity mapping: $W^{(0)} = X$ or
\[
W_\textrm{all}^{(0)} = \begin{bmatrix}
X_{\cdot 1} & X_{\cdot 2} & \dots & X_{\cdot D} \\ 
    \mathbf{1}_n &  \mathbf{0}_n & \dots & \mathbf{0}_n \\
    \vdots & & \ddots & \\ 
    \mathbf{0}_n &  \mathbf{0}_n & \dots & \mathbf{1}_n \\
\end{bmatrix},
\]
where $X_{\cdot d}$ denotes the $d^\textrm{th}$ column of $X$.
We use a prior mean of zero on the inner layer, although 
our software also supports an identity prior mean with the corresponding
derivatives (i.e., the columns of $W_\textrm{all}^{(0)}$ above).

\added{We task each surrogate with predicting the response and the gradient.}
We consider performance in root mean squared
error (RMSE, lower is better) which measures predictive accuracy, and
continuous ranked probability score 
\citep[CRPS, lower is better,][]{gneiting2007strictly} which incorporates
uncertainty quantification.\footnote{Technically, CRPS relies 
on a Gaussian posterior, but we use it here with DGPs given their 
conditional Gaussianity.}  For gradient predictions, partial derivatives 
are predicted independently, then performance metrics are averaged 
across $d=1,\dots,D$.  \added{Altogether, there are three key products of this work: 
gradient-enhanced DGP predictions of the response, gradient-enhanced DGP predictions of
the gradient, and non-enhanced DGP predictions of the gradient.  In the
following figures, these contributions are represented by the purple boxplots
(the geDGP response and gradient predictions) and the blue boxplots in the 
right panels (the DGP gradient predictions).}

\begin{figure}[ht!]
\centering
\includegraphics[width=\textwidth]{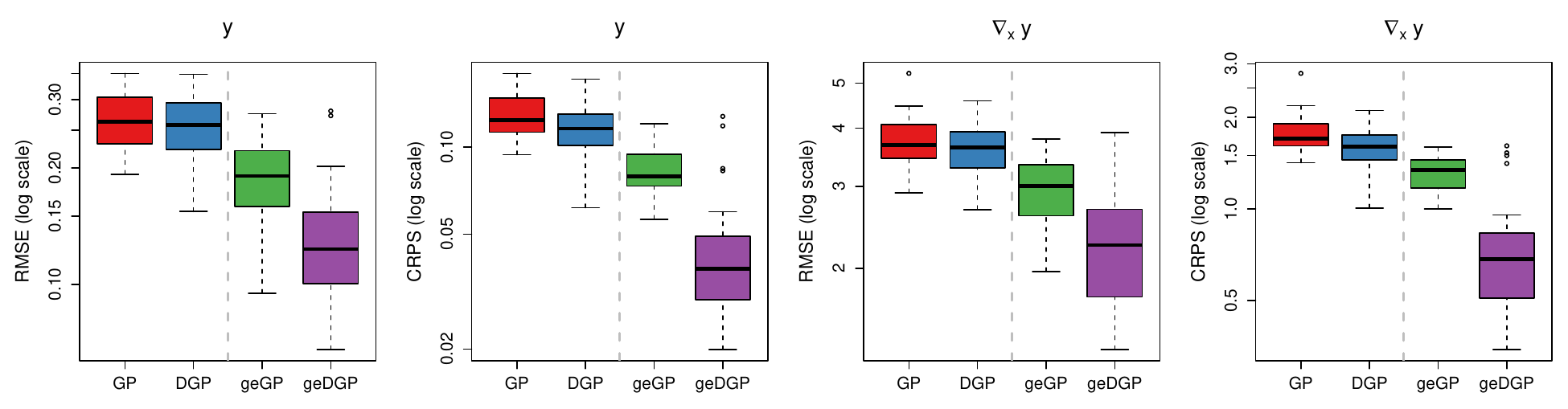}
\caption{Simulation results for the squiggle function with $D=2$ and $n=25$.}
\label{fig:squiggle_results}
\end{figure}

First, we consider the 2d ``squiggle'' function
\citep{duqling}, which featured previously in Figure~\ref{fig:squiggle},
with a training data size of $n=25$ \added{(meaning gradient-enhanced surrogates
use $N=75$)}.  \citet{rumsey2025all} found DGPs
consistently outperformed stationary GP alternatives on this function,
being better equipped to handle the flat regions and the ``S''-shaped
peak.  Figure~\ref{fig:squiggle_results} shows the performance of each
surrogate in predicting the response (left two panels) and its
gradient (right two panels) across the 30 repetitions.  Throughout, 
RMSE and CRPS are shown on the log scale.  The gray
dotted lines separate surrogates that are trained only on
$\{X, \mathbf{y}\}$ (left of the line) from the gradient-enhanced ones that
are trained on $\{X, \mathbf{y}, \nabla_x\mathbf{y}\}$ (right of the line).
The DGP performs a bit better than the stationary GP in predicting the response
and its gradient. Naturally, gradient-enhancement greatly improves the performance of both
surrogates.  The geDGP is far superior to the geGP on this nonstationary
function, with lower RMSE and CRPS across the board.  

\begin{figure}[ht!]
\centering
\includegraphics[width=\textwidth]{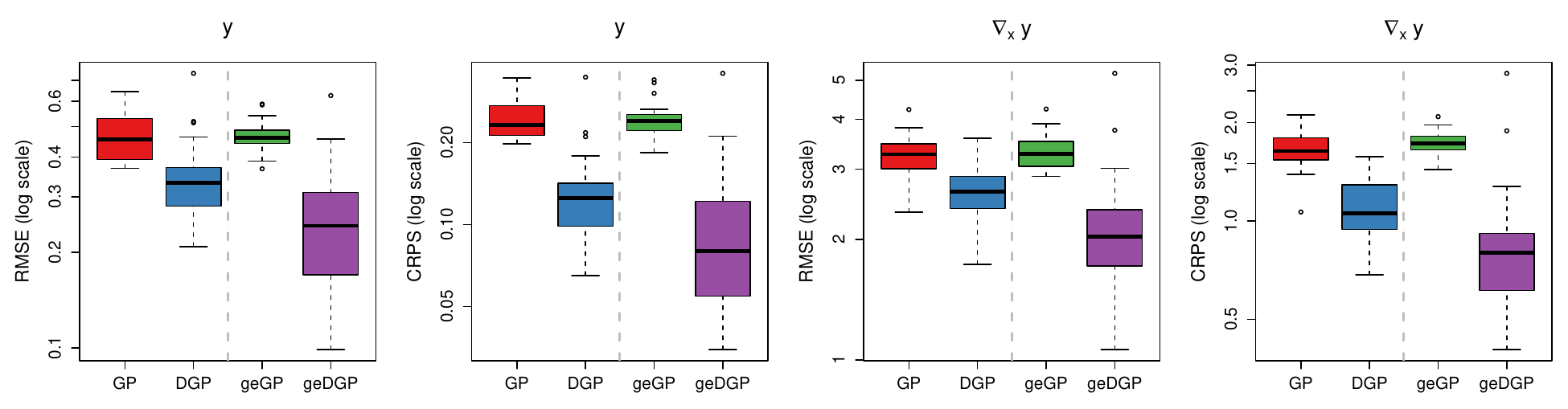}
\caption{Simulation results for the plateau function with $D=3$ and $n=30$.}
\label{fig:plateau_results}
\end{figure}

Next, we consider the ``plateau'' function from \citet{booth2025contour}
with $D=3$ and $n=30$ \added{($N=120$)}.  This function is characterized by flat regions with
a steep sloping drop between them.  Results are shown in 
Figure~\ref{fig:plateau_results}.  The improvement of the DGP over the GP
is starker here.  In this case, the DGP without gradient-enhancement outperforms
the GP with gradient-enhancement.  We take this as a strong indication of
the nonstationary nature of this function.  The geDGP consistently outperformed
the DGP.  Although there were occasional outliers in the geDGP's predictions of 
$\nabla_x\mathbf{y}$, these runs still had comparable performance in their 
predictions of $\mathbf{y}$.

\begin{figure}[ht!]
\centering
\includegraphics[width=\textwidth]{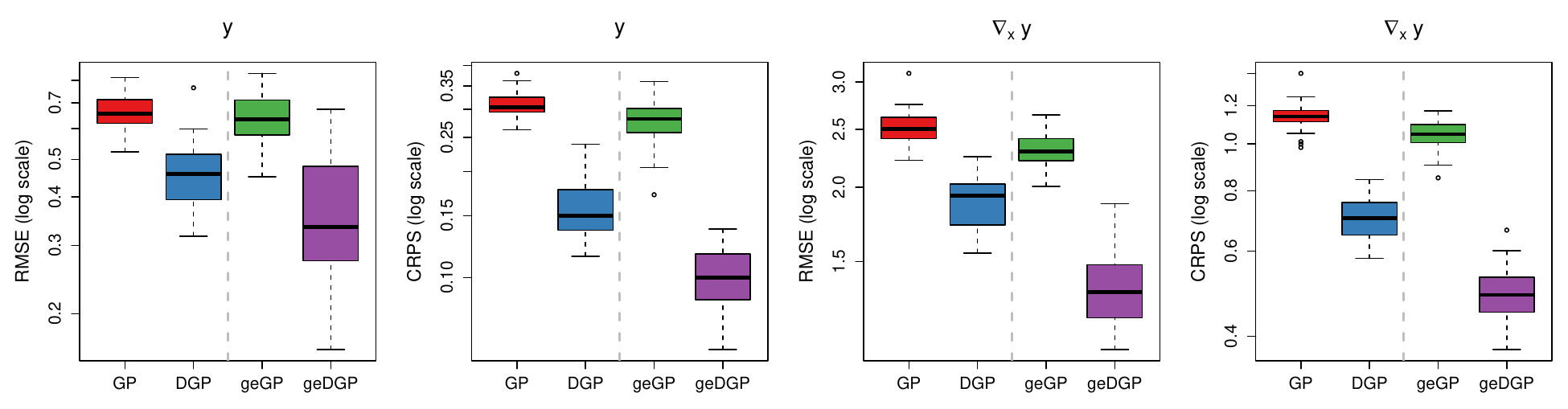}
\caption{Simulation results for the ignition function with $d=6$ and $n=100$
using Vecchia approximation.}
\label{fig:ignition_results}
\end{figure}

Finally, we consider the mock ``ignition'' function \citep{duqling,rumsey2025all}.  
This function mimics the yield of a fusion reaction as described in 
\citet{hatfield2019using} and exhibits nonstationarity due to a steep
``ignition cliff.'' We consider 6 inputs with $n=100$.  \added{Vecchia
approximation is advisable to accommodate the gradient-enhanced
surrogates with $N=700$; we use it for all four
surrogates to maintain consistency.}
Results are shown in Figure~\ref{fig:ignition_results}.  Again, we see
the nonstationary flexibility of the DGP can be more impactful than
gradient-enhancement with a stationary GP.  It is particularly noteworthy 
that the DGP is able to predict the gradient better than the 
gradient-enhanced GP.  \added{Unsurprisingly, the geDGP has the edge
across the board.}

\begin{figure}[ht!]
\centering
\includegraphics[width=\textwidth]{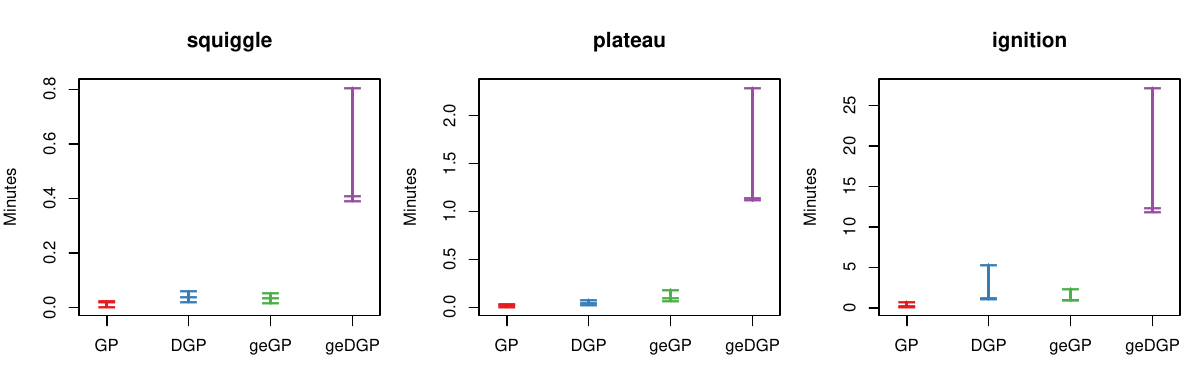}
\caption{\added{Computation time for generating 1,000 posterior samples
of all latent quantities (lower bar) and predictions of the
response (middle bar) and gradient (upper bar) 
across 100 of these samples.}}
\label{fig:timing}
\end{figure}

\added{For additional context, the lowest horizontal bars in 
Figure \ref{fig:timing} show the computation time required to generate 
1,000 MCMC samples of all latent 
quantities for a single repetition of each surrogate on each test function.
As expected, computation is heftier for the DGPs which require ESS,
and it grows with $D$, $n$, and the addition of gradient-enhancement.  
The middle horizontal bars indicate the additional
time required to obtain predictions of the response from each surrogate,
using $n_p = 100D$, aggregated across 100 of these samples.  In most 
situations, this
time is negligible compared to the time of MCMC training (the middle bar
often sits atop the lower bar).  The upper horizontal bars
show the additional time required to predict the gradient.  
Predicting all $D$-many partial derivatives is often the 
most computationally demanding feat.  All times were measured on a
MacBook Pro with an Apple M4 Max chip, 36 GB memory, and 14 CPU.}

%%%%%%%%%%%%%%%%%%%%%%%%%%%%%%%%%%%%%%%%%%%%%%%%%%%%%%%%%%%%%%%%%%%%%
\section{\added{Application to Quantum Mechanics}}\label{sec:quantum}
%%%%%%%%%%%%%%%%%%%%%%%%%%%%%%%%%%%%%%%%%%%%%%%%%%%%%%%%%%%%%%%%%%%%%

\added{Computer experiments are very useful in the field of quantum mechanics where
they are used to simulate the behavior of matter \citep{georgescu2014quantum}.
We are particularly interested in predicting the potential energy and forces
of a molecule as a function of its atomic structure (specifically the relative
positions of its atoms).  Quantum simulations
are very complex and can be computationally expensive.  In light of the sheer
number of unique molecules available, it is infeasible to obtain
more than a handful of observations for each molecule.  Large
efforts have gone into creating repositories of quantum simulation
data for general use \citep{gabellini2024openqdc}, yet even these massive
repositories contain limited evaluations of each molecule.}

\added{Here, we work with the SPICE data set as described in \citet{eastman2023spice}.
This data set contains observations of quantum chemistry simulations that return the 
potential energy and nuclear forces of a molecule as a function of that molecule's 
atomic positions. Crucially, the nuclear forces are the negative partial derivatives 
of the potential energy. We obtain and pre-process the data according to the 
git repository referenced in 
\citet{eastman2023spice}.}\footnote{\url{https://github.com/openmm/spice-dataset}}
\added{Each molecule is formally defined by its ``simplified molecular input line
entry system'' \citep[SMILE;][]{weininger1988smiles}, but is given a more
succinct type label for easy reference.  Here, we restrict our exercise
to four specific molecules, as listed in Table \ref{tab:molecules}.  The
first three molecules are made up of two atoms.  The position of each atom
is defined by its three dimensional coordinates, but since one atom is the 
``anchor point,'' the input variable is condensed into the 3 dimensional 
position of the second atom.  The fourth molecule contains 3 atoms, but two
of them are bonded hydrogen which do not change position relative to each other.
With both hydrogens serving as the anchor point, the relevant input is again
condensed into the 3 dimensional position of the final atom.  The SPICE data 
contains 11 observations of each 2-atom molecule and 25 observations of the 3-atom
molecule.}

\begin{center}
{\renewcommand*\arraystretch{1.1}
\begin{table}[h!]
\begin{tabular}{cccc}
\added{Type Label} & \added{SMILE} & \added{Common Name} & \added{Number of observations} \\
\hline
\added{0-13} & \added{[K+:2].[Br-:1]} & \added{potassium bromide} & \added{11} \\
\added{11-19} & \added{[Na+:2].[I-:1]} & \added{sodium iodide} & \added{11} \\
\added{6-14} & \added{[Li+:2].[Cl-:1]} & \added{lithium chloride} & \added{11} \\
\added{10-10-19} & \added{[H:1][H:2].[Na+:3]} & \added{N/A} & \added{25} \\
\end{tabular}
\caption{\added{Molecules considered from the SPICE data.}}
\label{tab:molecules}
\end{table}
}
\end{center}
\vspace{-1cm}

\added{We reserve about two thirds of the data for training and one third of the data
for testing (a 7-4 split for the 2-atom molecules and a 16-9 split for the 3-atom molecule).
We consider GP and DGP surrogates both with and without gradient enhancement,
as described in Section \ref{sec:results}.  Gradient-enhanced models are
trained on observations of the potential energy and the corresponding nuclear forces.
Each model is tasked with predicting both the energy ($y$) and forces ($\nabla_x y$) given
the atomic positions of the testing observations.  In real-data scenarios, we recommend
assessing MCMC trace plots for convergence to inform burn-in and 
thinning choices.  We investigate these for a single training/testing split of the 0-13
molecule in Supplement \ref{supp:trace}, but we find the defaults provided in
Section \ref{sec:results} work well here too.}

\begin{figure}[ht!]
\centering
\includegraphics[width=\textwidth]{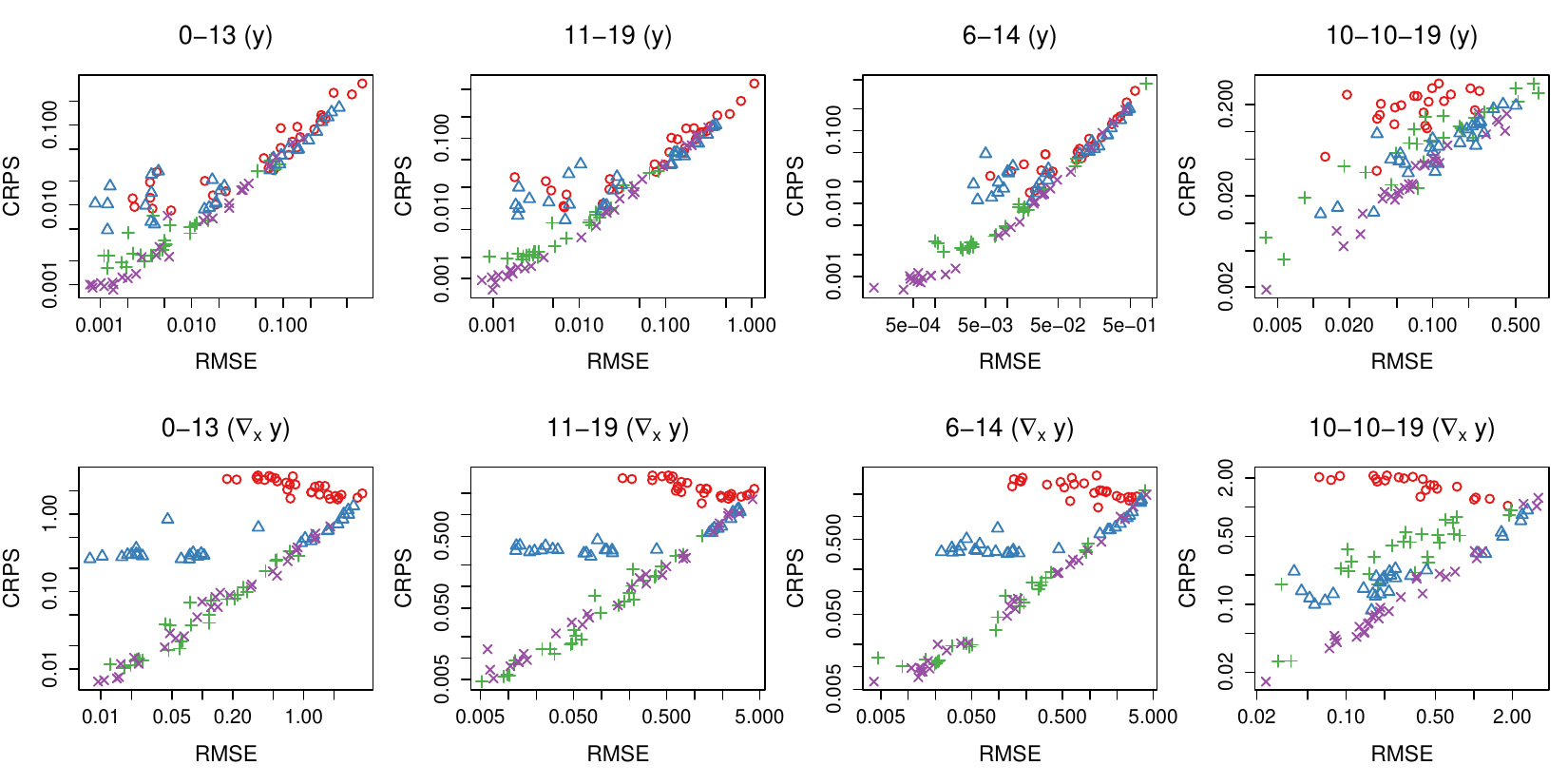}\\
\includegraphics[width=0.4\textwidth]{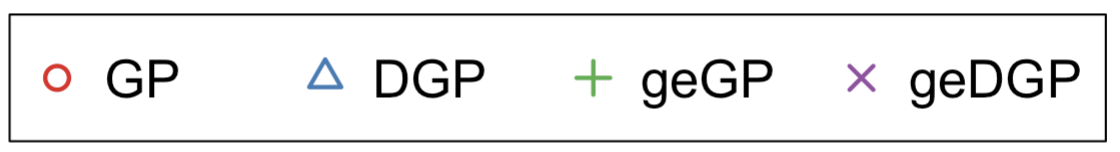}
\caption{\added{Surrogate performance in predicting the energy (top row) and forces
(bottom row, averaged across dimension) for 4 molecules from the SPICE data.}}
\label{fig:quantum}
\end{figure}

\added{Figure \ref{fig:quantum} displays
performance across 30 different training/testing splits.  Results in the lower
left corner (with the smallest RMSE and smallest CRPS) are preferred.
The traditional DGP offers much better performance than the stationary GP, particularly
in predicting the gradient, which indicates there is significant nonstationary 
in the response surface.  Both gradient-enhanced variations outperform their 
non-enhanced counterparts.  As expected, conditioning on gradient information 
improves predictions of the response and the gradient.  Notably, the gradient-enhanced
DGP outperforms across the board, combining the nonstationarity of the DGP with the
valuable information gradients provide.  These results suggest that geDGP surrogates
can be valuable tools to the field of quantum mechanics; extension to more complex
molecules is an interesting avenue for future work.}

%%%%%%%%%%%%%%%%%%%%%%%%%%%%%%%%%%%%%%%%%%%%%%%%%%%%%%%%%%%%%%%%%%%%%
\section{Discussion}\label{sec:discuss}
%%%%%%%%%%%%%%%%%%%%%%%%%%%%%%%%%%%%%%%%%%%%%%%%%%%%%%%%%%%%%%%%%%%%%

We have provided an upgraded Bayesian DGP modeling framework for 
surrogate modeling of deterministic computer experiments that 
enables both DGP gradient predictions and gradient-enhanced DGPs.  
Our fully-Bayesian implementation, which leverages elliptical slice 
sampling of latent nodes and their gradients in conjunction with the 
multivariate chain rule, performs well on a variety of 
nonstationary functions \added{and real quantum mechanics computer
experiments, offering improvements in accuracy and UQ
over geGPs and non-enhanced DGPs.  Although our focus is on 
expensive computer experiments where
training data sizes are severely limited, gradient-enhanced DGPs may hit
computational bottlenecks even for moderate $n$ (as in our ignition function
example where $N=700$).  In those cases, the incorporation of 
optional Vecchia approximation can circumvent cubic computational costs.}
All methods are open-source and publicly available in the 
{\tt deepgp} package on CRAN \citep{deepgp}.

\added{In this work, we focused exclusively on deterministic experiments 
with noise-free response and gradient observations.  A notable avenue for future 
work is the extension of our methods to
stochastic computer experiments.  Our software does support gradient predictions
from a non-enhanced DGP with an estimated nugget parameter, but the current
implementation assumes the response and gradient share the same nugget, 
which may be an inappropriate assumption.  Gradient-enhancement is a whole
different story.  If response observations are noisy, it is safe to 
assume any observed gradients will also be noisy, possibly to differing
degrees.  Incorporating noise
estimation for the response and gradients will surely bring additional
challenges and will require careful development.  Even in the deterministic
case, selection of the jitter vaue and thorough comparisons against
alternative methods to ensure numerical stability \citep[like the random 
feature expansion method of][]{hung2021random} is deserving of further
investigation.}

\added{Missing gradient information is another interesting direction for
gradient-enhanced DGPs.  It is possible that some runs of a computer experiment
may not provide gradient observations, while others will (thanks to numerical
or convergence issues inside the simulator).  Our current implementation
is all-or-nothing when it comes to gradient-enhancement, but it could be
upgraded to accommodate missing gradients.}

\added{An additional direction for future work is in refinement of the 
Vecchia approximation ordering and conditioning for gradient-enhanced models.  
To our knowledge, this work is the first to consider Vecchia approximation for 
geGPs (and of course geDGPs).  The unique structure of $\mathbf{y}$ and 
$\nabla_x\mathbf{y}$ (with many ``ties''
in pairwise distances) suggests that some ordering and conditioning 
set options may be better than others.  While we proposed choices
that seem sensible and worked well in our exercises, they were rather heuristic.
A thorough investigation of optimal ordering/conditioning for gradient-enhanced
Vecchia-approximated GP and DGP surrogates is warranted.}

\subsection*{Supplementary Material}

\added{Supplementary Material includes: derivations of the gradient for the Gaussian
and Mat\`{e}rn kernels, formulaic details of the benchmark functions used in 
Section \ref{sec:results}, comparisons of existing DGP and geGP methodologies
on the benchmark functions, and trace plots for surrogate fits to the
SPICE data.}

\subsection*{Acknowledgements}

This work was supported by the U.S. National Science Foundation under 
Award Number 2533443.

\singlespacing
\bibliographystyle{ba}
\bibliography{main}

\newpage
\appendix
\begin{center}
\Large{SUPPLEMENTARY MATERIAL}
\end{center}

\section{Gradients of the Gaussian Kernel}\label{supp:gaussian}
 
The Gaussian (or squared exponential) kernel, defined as
\[
K(\mathbf{x}_i, \mathbf{x}_j) = \mathrm{exp}\left(-\sum_{p=1}^D
\frac{(\mathbf{x}_{ip} - \mathbf{x}_{jp})^2}{\theta_p}\right),
\]
has the following derivatives (Eq.~\ref{eq:kernel}):
\[
\begin{aligned}
K_{d0}(\mathbf{x}_i, \mathbf{x}_j) &= \frac{\partial}{\partial x_i^d}K(\mathbf{x}_i, \mathbf{x}_j) \\
  &= \frac{\partial}{\partial x_i^d}\left[\mathrm{exp}\left(-\sum_{p=1}^D\frac{(x_{ip} - x_{jp})^2}{\theta_p}\right)\right] \\
    &= K(\mathbf{x}_i, \mathbf{x}_j) * \frac{-2}{\theta_d} * \left(x_{id} - x_{jd}\right) \\
K_{0d}(\mathbf{x}_i, \mathbf{x}_j) &= \frac{\partial}{\partial x_j^d} K(\mathbf{x}_i, \mathbf{x}_j) \\
  &= \frac{\partial}{\partial x_j^d}\left[\mathrm{exp}\left(-\sum_{p=1}^D\frac{(x_{ip} - x_{jp})^2}{\theta_p}\right)\right] \\
  &= K(\mathbf{x}_i, \mathbf{x}_j) * \frac{2}{\theta_d}*\left(x_{id} - x_{jd}\right) \\
K_{dd}(\mathbf{x}_i, \mathbf{x}_j) &= \frac{\partial}{\partial x_i^d}\left[\frac{\partial}{\partial x_j^d} K(\mathbf{x}_i, \mathbf{x}_j)\right] \\
  &= \frac{\partial}{\partial x_i^d}\left[K(\mathbf{x}_i, \mathbf{x}_j) * \frac{2}{\theta_d}*\left(x_{id} - x_{jd}\right)\right] \\
  &= \frac{2}{\theta_d}\times K(\mathbf{x}_i, \mathbf{x}_j) - \frac{4}{\theta_d^2}(x_{id}-x_{jd})^2 * K(\mathbf{x}_i, \mathbf{x}_j) \\
  &= \frac{2}{\theta_d}\left(1 - \frac{2}{\theta_d}(x_i^d-x_j^d)^2\right)*K(\mathbf{x}_i, \mathbf{x}_j) \\
K_{df}(\mathbf{x}_i, \mathbf{x}_j) &= \frac{\partial}{\partial x_i^d}\left[\frac{\partial}{\partial x_j^f}K(\mathbf{x}_i, \mathbf{x}_j)\right] \\
  &= \frac{\partial}{\partial x_i^d}\left[K(\mathbf{x}_i, \mathbf{x}_j) * \frac{2}{\theta_f}*\left(x_{if} - x_{jf}\right)\right] \\
  &= \frac{2}{\theta_f}(x_{if}-x_{jf}) * K(\mathbf{x}_i, \mathbf{x}_j) * \frac{-2}{\theta_d} * \left(x_{id} - x_{jd} \right) \\
  &= \frac{-4}{\theta_f\theta_d}(x_{if}-x_{jf})(x_{id}-x_{jd}) * K(\mathbf{x}_i, \mathbf{x}_j) \\
\end{aligned}
\]

\section{\added{Gradients of the Mat\`{e}rn-$5/2$ Kernel}}\label{supp:matern}
 
\added{The Mat\`{e}rn-$5/2$ kernel, defined as
\[
K(\mathbf{x}_i, \mathbf{x}_j) = \left(1 + r + \frac{1}{3}r^2\right)\mathrm{exp}\left(-r\right)
\quad\textrm{for}\quad
r = \sqrt{5\sum_{p=1}^D \frac{(\mathbf{x}_{ip} - \mathbf{x}_{jp})^2}{\theta_p}},
\]
has the following derivatives (Eq.~\ref{eq:kernel}):
\[
\begin{aligned}
K_{d0}(\mathbf{x}_i, \mathbf{x}_j) &= \frac{\partial}{\partial x_i^d}K(\mathbf{x}_i, \mathbf{x}_j) \\
  &= \frac{\partial}{\partial x_i^d}\left[\left(1 + r + \frac{1}{3}r^2\right)\mathrm{exp}(-r)\right] \\
    &= \frac{-5}{3\theta_d}(x_{id}-x_{jd})(1 + r)\mathrm{exp}(-r)\\
K_{0d}(\mathbf{x}_i, \mathbf{x}_j) &= \frac{\partial}{\partial x_j^d} K(\mathbf{x}_i, \mathbf{x}_j) \\
  &= \frac{\partial}{\partial x_j^d}\left[\left(1 + r + \frac{1}{3}r^2\right)\mathrm{exp}(-r)\right] \\
  &= \frac{5}{3\theta_d}(x_{id}-x_{jd})(1 + r)\mathrm{exp}(-r)\\
K_{dd}(\mathbf{x}_i, \mathbf{x}_j) &= \frac{\partial}{\partial x_i^d}\left[\frac{\partial}{\partial x_j^d} K(\mathbf{x}_i, \mathbf{x}_j)\right] \\
  &= \frac{\partial}{\partial x_i^d}\left[\frac{5}{3\theta_d}(x_{id}-x_{jd})(1+r)\mathrm{exp}(-r)\right] \\
  &= \frac{5}{3\theta_d}\mathrm{exp}(-r)\left(1 + r -\frac{5}{\theta_d}(x_{id}-x_{jd})^2\right) \\
K_{df}(\mathbf{x}_i, \mathbf{x}_j) &= \frac{\partial}{\partial x_i^d}\left[\frac{\partial}{\partial x_j^f}K(\mathbf{x}_i, \mathbf{x}_j)\right] \\
  &= \frac{\partial}{\partial x_i^d}\left[\frac{5}{3\theta_f}(x_{if}-x_{jf})(1+r)\mathrm{exp}(-r)\right] \\
  &= \frac{-25}{3\theta_f\theta_d}(x_{if}-x_{jf})(x_{id}-x_{jd})\mathrm{exp}(-r)
\end{aligned}
\]}

\section{Benchmark Functions}\label{supp:functions}

\subsection*{Squiggle Function} For $x_1,x_2\in [0, 1]$ the ``squiggle'' function 
\citep{duqling} is defined as
\[
f(\mathbf{x}) =  \frac{1}{\sqrt{2\pi\sigma^2}} * x_1 * x_2 *
    \mathrm{exp}\left(-\frac{1}{2\sigma^2} (x_2 - \mu)^2\right)
    \quad\textrm{where}\quad
    \mu = \frac{1}{4}\sin(2\pi x_1^2) - \frac{1}{10}x_1 + \frac{1}{2}.
\]
Its partial derivatives are
\[
\begin{aligned}
\frac{\partial f}{\partial x^1} 
    &= \frac{1}{\sqrt{2\pi\sigma^2}} * x_1 * x_2 * \mathrm{exp}\left(-\frac{1}{2\sigma^2} 
    (x_2 - \mu)^2\right) * \frac{1}{\sigma^2}(x_2 - \mu)*\frac{\partial\mu}{\partial x_1}
    + \\
    & \quad \frac{1}{\sqrt{2\pi\sigma^2}} * x_2 * \mathrm{exp}\left(-\frac{1}{2\sigma^2} 
    (x_2 - \mu)^2\right) \\
\frac{\partial f}{\partial x^2} 
    &= \frac{1}{\sqrt{2\pi\sigma^2}} * x_1 * x_2 * \mathrm{exp}\left(-\frac{1}{2\sigma^2} 
    (x_2 - \mu)^2\right) * \frac{1}{\sigma^2}(\mu - x_2)
    + \\
    & \quad \frac{1}{\sqrt{2\pi\sigma^2}} * x_1 * \mathrm{exp}\left(-\frac{1}{2\sigma^2} 
    (x_2 - \mu)^2\right) \\
\end{aligned}
\]
where
\[
\frac{\partial\mu}{\partial x_1} = \pi x_1\cos(2\pi x_1^2) - \frac{1}{10}.
\]

\subsection*{Plateau Function} For $x_i\in[-2, 2]$, $\,i=1,\dots,D$, the ``plateau'' 
function \citep{booth2025contour} is defined as
\[
f(\mathbf{x}) = 2*\Phi\left(\sqrt{2}\left(-4-3\sum_{i=1}^D x_i\right)\right) - 1,
\]
where $\Phi$ is the standard Gaussian CDF.  Its partial derivatives are
\[
\frac{\partial f}{\partial x^d} = -24\sqrt{2}*\phi
\left(\sqrt{2}\left(-4-3\sum_{i=1}^D x_i\right)\right)
\quad\textrm{for}\quad i=1,\dots,D,
\]
where $\phi$ is the standard Gaussian PDF.

\subsection*{Ignition Function}

For $x_i\in[0,1]$, $\, i=1,\dots,D$, the mock ``ignition'' function
\citep{duqling} is defined as
\[
\begin{aligned}
f(\mathbf{x}) &= \log_{10}\left(q\right)\quad\textrm{where} \\
q &= r^5\left(1 + 200000t\right) \\
t &= \Phi\left(10\sqrt{2}*(r-2)\right) \\ 
r &= \sqrt{\sum_{i=1}^D x_i^2}
\end{aligned}
\]
Its partial derivatives are
\[
\begin{aligned}
\frac{\partial f}{\partial x^d} &= \frac{dq}{q\log(10)}\quad\textrm{where}\\
dq &= r^5\left(200000*dt\right) + \left(1 + 200000t\right)*5*r^4*dr \\
dt &= 10\sqrt{2}*\phi\left(10\sqrt{2}*(r-2)\right)*dr \\
dr &= \frac{x_{ij}}{r}
\quad\quad\textrm{for}\quad i=1,\dots,D.
\end{aligned}
\]

\section{\added{Comparing existing surrogate models}}\label{supp:existing}

\added{In this section, we consider a variety of existing surrogate models
on the three synthetic functions discussed in Section \ref{sec:results}.  To keep
the results presented there from getting too cluttered, we included only the best
performers from those considered here.}

\begin{figure}[ht!]
\centering
\includegraphics[width=0.9\textwidth]{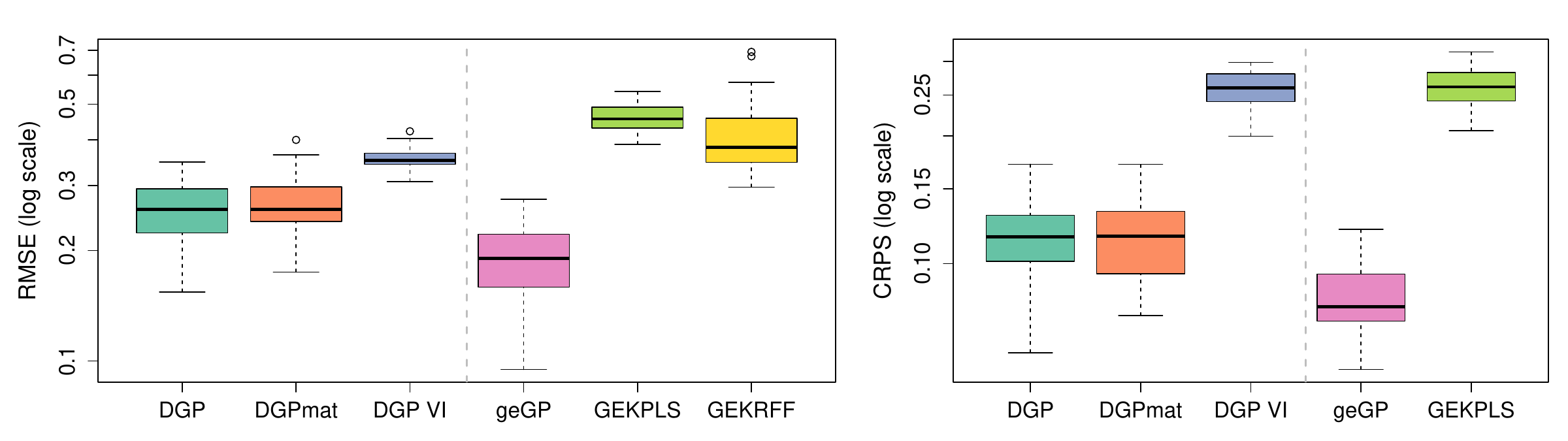}\\
\caption{\added{Performance of existing DGP and gradient-enhanced GP models on the squiggle function.}}
\label{fig:squiggle_existing}
\end{figure}

\begin{figure}[ht!]
\centering
\includegraphics[width=0.9\textwidth]{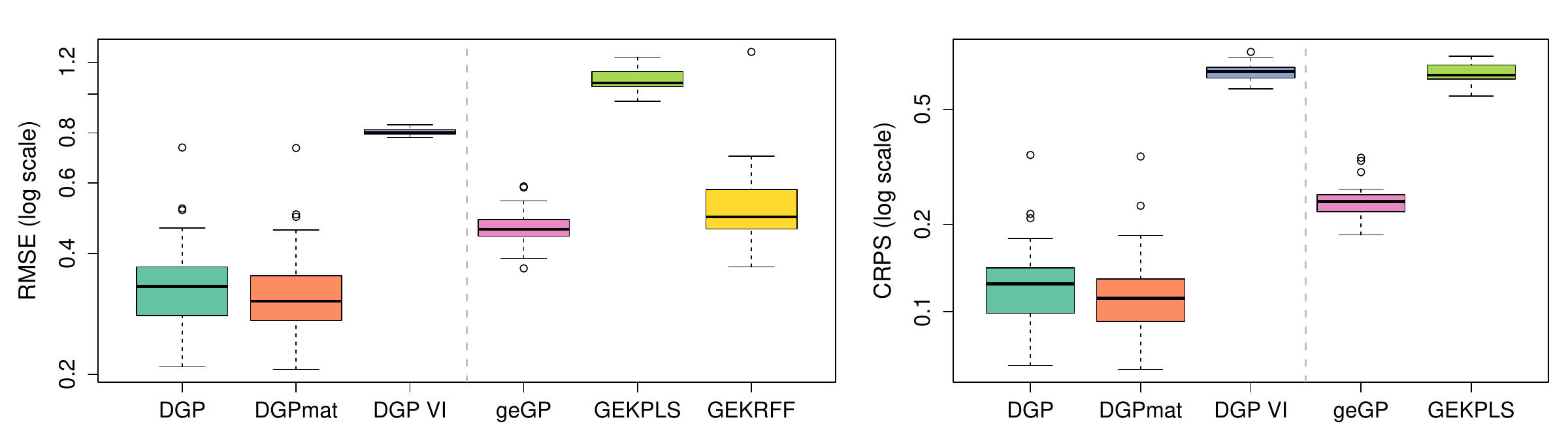}\\
\caption{\added{Performance of existing DGP and gradient-enhanced GP models on the plateau function.}}
\label{fig:plateau_existing}
\end{figure}

\begin{figure}[ht!]
\centering
\includegraphics[width=0.9\textwidth]{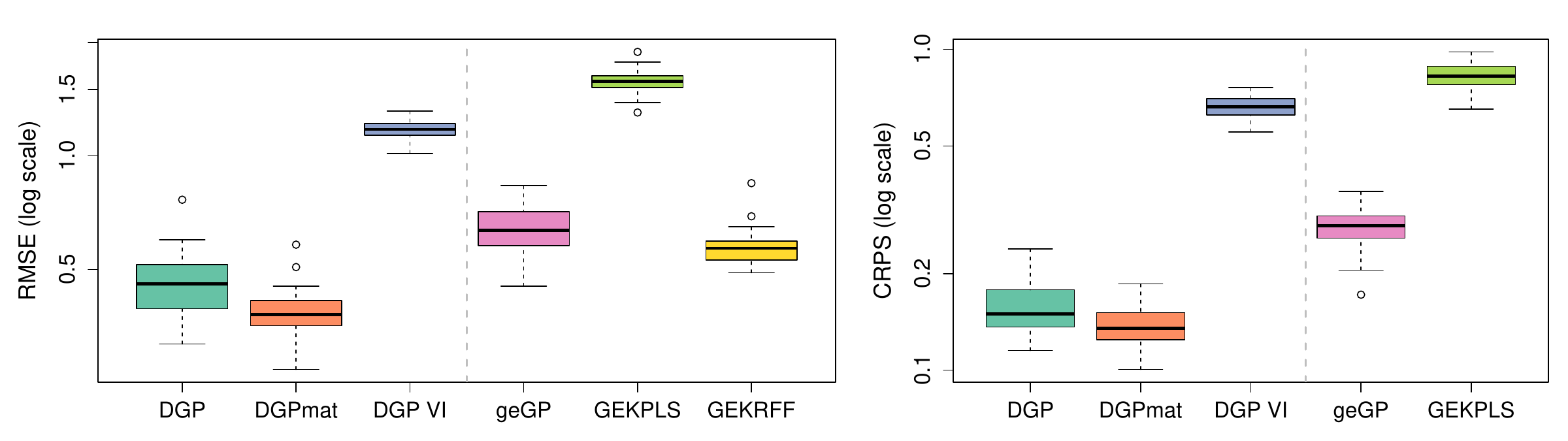}\\
\caption{\added{Performance of existing DGP and gradient-enhanced GP models on the ignition function.}}
\label{fig:ignition_existing}
\end{figure}

\added{First, we consider three variations of a two-layer deep Gaussian process:}
\begin{itemize}
    \item \added{DGP: a two-layer DGP from the {\tt deepgp} R package \citep{deepgp}
    with the squared exponential kernel.  Uses elliptical slice sampling of the 
    latent layer and Metropolis Hastings sampling of lengthscales 
    following \citet{sauer2023active}.  Uses Vecchia approximation for the ignition function
    \citep{sauer2023vecchia}.}
    \item \added{DGPmat: same as DGP, but with the Mat\`{e}rn-$5/2$ kernel.}
    \item \added{DGP VI: a two-layer DGP from the {\tt gpflux} python package
    \citep{gpflux} with the squared exponential kernel.  Uses ``doubly stochastic variational inference'' with inducing point approximations 
    \citep{salimbeni2017doubly}.  We followed
    the defaults provided in the software's documentation without additional manual tuning.}
\end{itemize}
\added{Second, we consider three variations of a gradient-enhanced Gaussian process:}
\begin{itemize}
    \item \added{geGP: a gradient-enhanced GP from the {\tt deepgp} R package \citep{deepgp}
    with the squared exponential kernel as described in this manuscript.  Uses Vecchia 
    approximation for the ignition function.}
    \item \added{GEKPLS: a gradient-enhanced GP using partial least squares \citep{bouhlel2019gradient}
    from the {\tt Surrogate Modeling Toolbox} in python \citep{smt}.}
    \item \added{GEKRFF: a gradient-enhanced GP using random Fourier features 
    following \citet{hung2021random}.  Implemented in MATLAB.  We obtained the MATLAB 
    code directly from the authors of this work.}
\end{itemize}
\added{All gradient-enhanced GP implementations use the Gaussian/squared exponential kernel for 
consistency (the {\tt deepgp} package is the only one that offers the Mat\`{e}rn-$5/2$ kernel).
Simulation settings match those described in Section \ref{sec:results}.  Root mean
squared error and continuous ranked probability score for predicting the response $y$ are
shown in Figures \ref{fig:squiggle_existing}--\ref{fig:ignition_existing}.  GEKRFF does not provide
variance estimates, so it is omitted from the CRPS results.  Boxplots show the
distribution of 30 repetitions.}

\added{Both fully-Bayesian DGP implementations (DGP and DGPmat) outperform the variational inference
implementation (DGP VI).  VI strategies are better suited for large data sets and struggle in these
smaller data settings.  Inducing point approximations can also be low fidelity when data is not
abundant; previous work has shown Vecchia approximation to be more promising \citep{sauer2023vecchia}.
The Mat\`{e}rn-$5/2$ kernel slightly improved DGP performance for the ignition function, but did
not offer an advantage on the other two functions.}

\added{The GEKPLS model struggled across the board.  Although the software is user-friendly, 
documented, and easy to use, the partial least squares approximation is designed for higher 
dimensional settings.  The GEKRFF model underperformed on the squiggle function, but was able
to match the performance of the geGP on the ignition function.  Nevertheless, the lack of uncertainty
quantification from GEKRFF makes the geGP implementation preferable in many settings.}

\section{\added{MCMC Convergence}}\label{supp:trace}

\begin{figure}[ht!]
\centering
\includegraphics[width=0.9\textwidth]{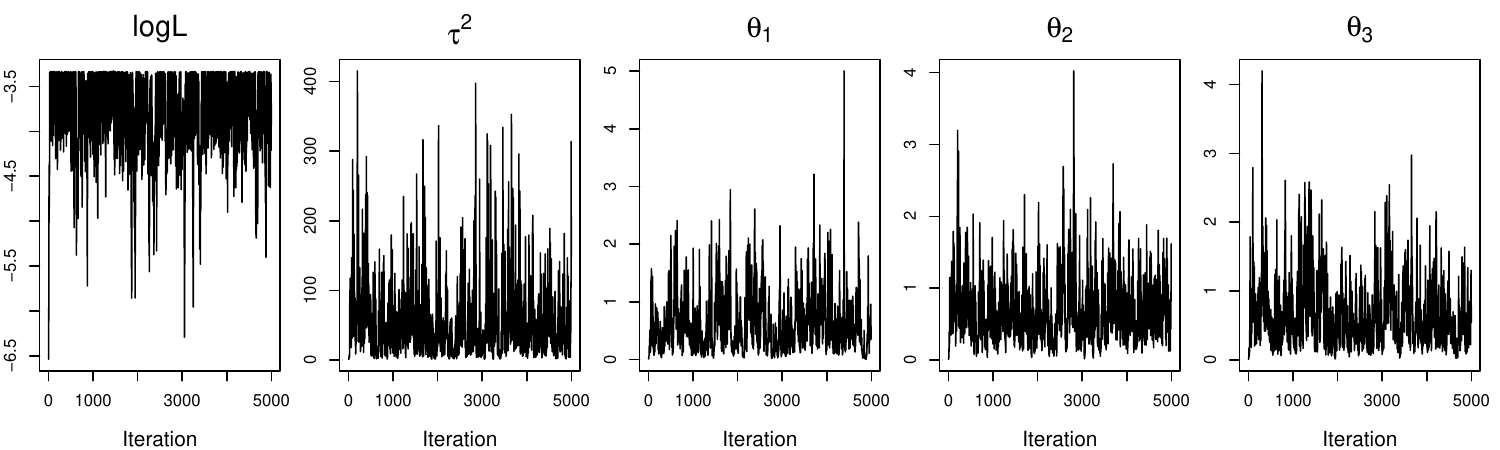}
\caption{\added{Trace plots for the GP fit to 7 observations of the 0-13 
molecule from the SPICE data.}}
\label{fig:trace_gp}
\end{figure}

\begin{figure}[ht!]
\centering
\includegraphics[width=0.9\textwidth]{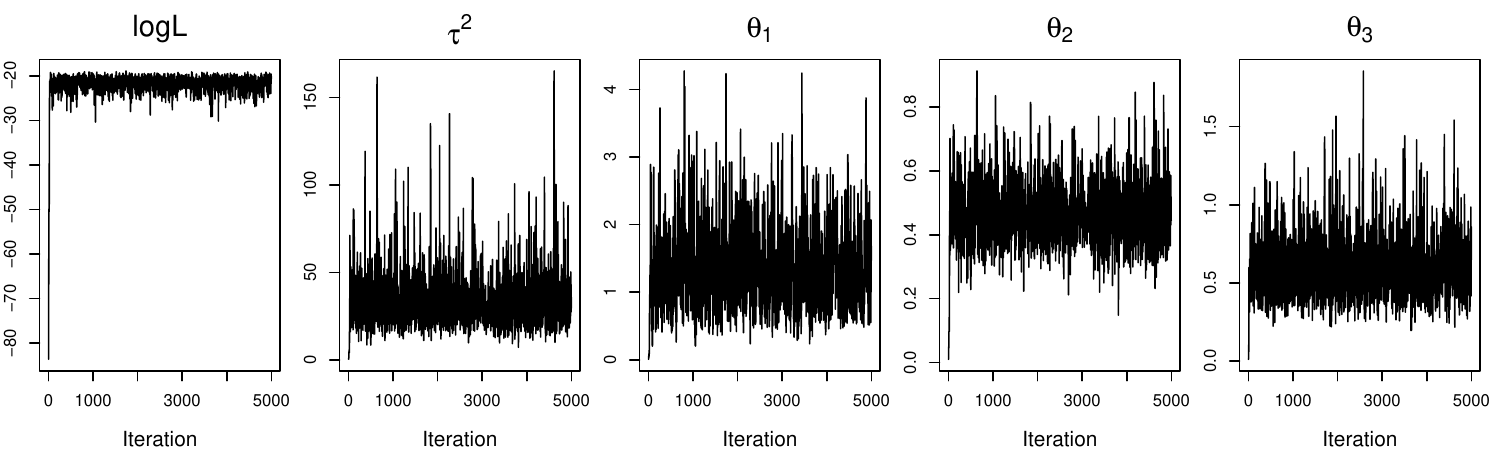}
\caption{\added{Trace plots for the geGP fit to 7 observations of the 0-13 
molecule from the SPICE data.}}
\label{fig:trace_gegp}
\end{figure}

\added{In this section, we present trace plots for the four surrogates
entertained for the SPICE data set in Section \ref{sec:quantum}.
We consider a single training/testing partition for the 0-13 molecule,
with 7 training observations in 3 dimensions.  We use the default priors
and initial values provided in the {\tt deepgp} package.  For the 
GP and geGP surrogate, the inferred quantities are the scale 
parameter $\tau^2$ and the separable lengthscale parameters 
$\boldsymbol\theta = [\theta_1, \theta_2, \theta_3]$.  We 
additionally include trace plots of the Gaussian log likelihoods 
(Eqs.~\ref{eq:gplogl} and \ref{eq:gegplogl}, respectively).
Trace plots for 5,000 samples of each of these are shown in 
Figures \ref{fig:trace_gp} and \ref{fig:trace_gegp}.  Visual
inspection indicates that MCMC convergence is reached very
quickly.}

\begin{figure}[ht!]
\centering
\includegraphics[width=\textwidth]{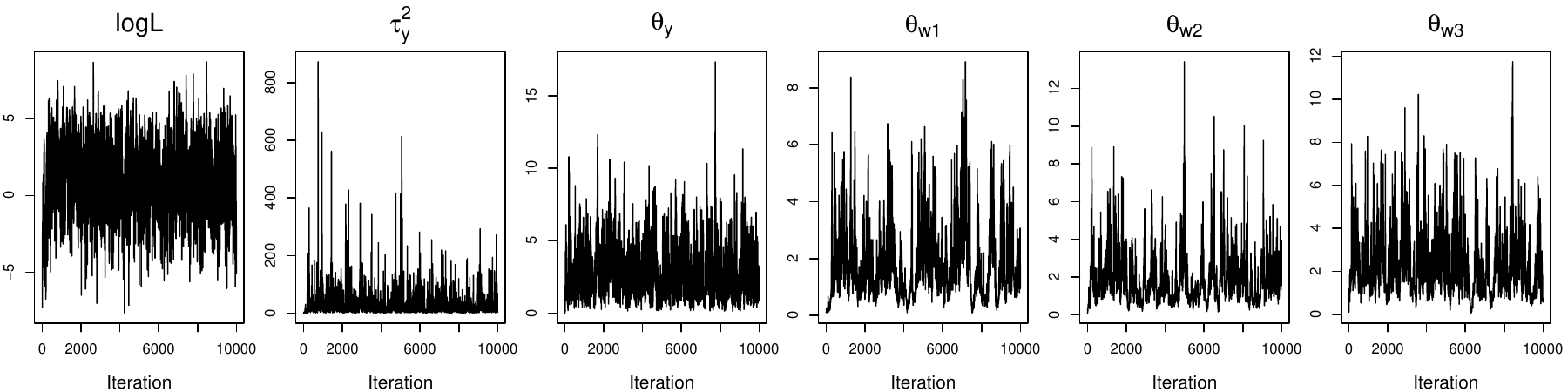}
\caption{\added{Trace plots for the DGP fit to 7 observations of the 0-13 
molecule from the SPICE data.}}
\label{fig:trace_dgp}
\end{figure}

\begin{figure}[ht!]
\centering
\includegraphics[width=\textwidth]{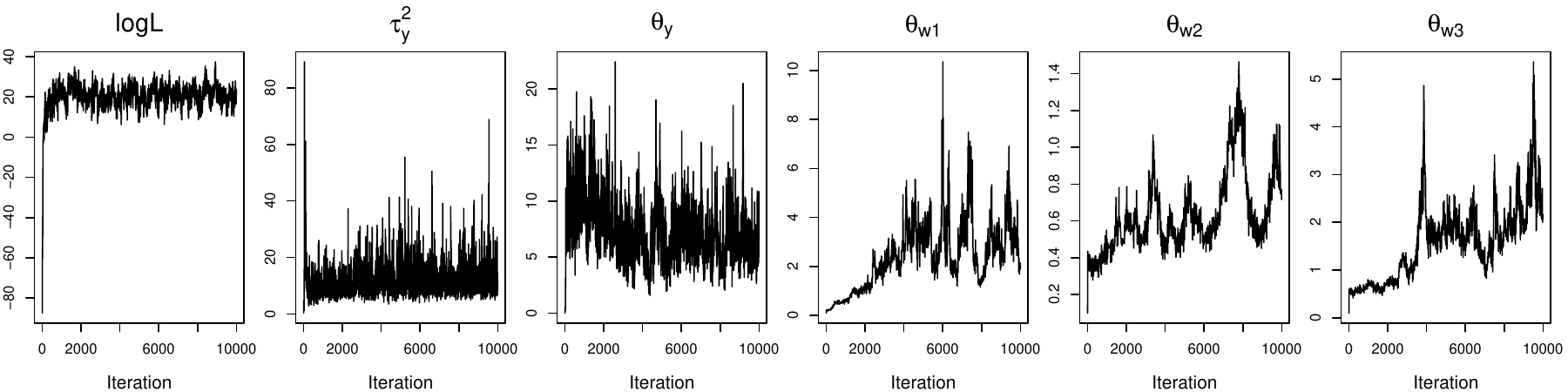}
\caption{\added{Trace plots for the geDGP fit to 7 observations of the 0-13 
molecule from the SPICE data.}}
\label{fig:trace_gedgp}
\end{figure}

\added{The DGP and geDGP fits to this data contains additional 
latent quantities.  The most notable is the latent $W$ (or $W_\textrm{all}$
for the geDGP) with 3 nodes, but there are also more
lengthscale parameters (with one on the outer layer and 3 on the
inner layer).  While we can investigate trace plots of the scalars easily,
the functional nature of $W$ makes its burn-in hard to visualize.  We
believe the best glimpse at convergence of $W$ comes from the trace plot
of the outer Gaussian log likelihood.  The outer likelihood will trend
upwards if ESS samples are still progressing but will settle in if samples
have generally converged.  Figures \ref{fig:trace_dgp} and \ref{fig:trace_gedgp}
show the trace plots for 10,000 samples of these fits.  While we include
the inner lengthscales $\theta_{w1}$, $\theta_{w2}$, and $\theta_{w3}$, we
are most concerned with the quantities associated with the outer GP layer
as it is our primary regression model.  The inner lengthscales affect the
prior draws of $W$ used in ESS proposals, but they are otherwise irrelevant.
We see quick and sufficient convergence from the outer log likelihood,
the scale $\tau^2_y$, and the outer lengthscale $\theta_y$ in both cases.}

\added{Retaining more MCMC samples to use for predictions will require more 
computation.  Although retained samples $t\in\mathcal{T}$ can be handled
in parallel, there is still a computational burden to retaining too many.
For this reason, we thinned the MCMC chains down to 1,000 samples for
all of our exercises.  We acknowledge that retaining more samples could
provide larger effective sample sizes and more robust predictions.} 

\end{document}